\documentclass[3p,authoryear]{elsarticle}

\journal{Icarus}

\usepackage{natbib}

\usepackage{amssymb}

\usepackage{lineno}

\usepackage{textcomp}

\usepackage{color}
\usepackage[usenames,dvipsnames]{xcolor}
\usepackage{ulem}

\usepackage{doi}

\usepackage{hyperref}
\hypersetup{colorlinks=true, urlcolor=blue}

\begin{document}

\begin{frontmatter}



\title{Laboratory Observations and Simulations of Phase Reddening\tnoteref{label1}\tnoteref{label2}}
\tnotetext[label1]{\doi{10.1016/j.icarus.2014.06.010}}
\tnotetext[label2]{\copyright 2017. This manuscript version is made available under the CC-BY-NC-ND 4.0 licence:\\ \url{https://creativecommons.org/licenses/by-nc-nd/4.0/}}


\author[DLR]{S.E.~Schr\"oder\corref{cor1}}
\author[UP]{Ye.~Grynko}
\author[UB]{A.~Pommerol}
\author[TUB]{H.U.~Keller}
\author[UB]{N.~Thomas}
\author[NA]{T.L.~Roush}

\cortext[cor1]{Corresponding author}

\address[DLR]{Deutsches Zentrum f\"ur Luft- und Raumfahrt (DLR), 12489 Berlin, Germany}
\address[UP]{Universit\"at Paderborn, 33098 Paderborn, Germany}
\address[UB]{Physikalisches Institut, Universit\"at Bern, 3012 Bern, Switzerland}
\address[TUB]{Institut f\"ur Geophysik und extraterrestrische Physik (IGEP), Technische Universit\"at Braunschweig, 38106 Braunschweig, Germany}
\address[NA]{NASA Ames Research Center, Moffett Field, CA 94035-0001, U.S.A.}

\begin{abstract}

The visible reflectance spectrum of many solar system bodies changes with changing viewing geometry for reasons not fully understood. It is often observed to redden (increasing spectral slope) with increasing solar phase angle, an effect known as phase reddening. Only once, in an observation of the Martian surface by the Viking~1 lander, was reddening observed up to a certain phase angle with bluing beyond, making the reflectance ratio as a function of phase angle shaped like an arch. However, in laboratory experiments this arch-shape is frequently encountered. To investigate this, we measured the bidirectional reflectance of particulate samples of several common rock types in the 400-1000~nm wavelength range and performed ray-tracing simulations. We confirm the occurrence of the arch for surfaces that are forward scattering, i.e.\ are composed of semi-transparent particles and are smooth on the scale of the particles, and for which the reflectance increases from the lower to the higher wavelength in the reflectance ratio. The arch shape is reproduced by the simulations, which assume a smooth surface. However, surface roughness on the scale of the particles, such as the \citet{HH63} fairy castles that can spontaneously form when sprinkling a fine powder, leads to monotonic reddening. A further consequence of this form of microscopic roughness (being indistinct without the use of a microscope) is a flattening of the disk function at visible wavelengths, i.e.\ Lommel-Seeliger-type scattering. The experiments further reveal monotonic reddening for reflectance ratios at near-IR wavelengths. The simulations fail to reproduce this particular reddening, and we suspect that it results from roughness on the surface of the particles. Given that the regolith of atmosphereless solar system bodies is composed of small particles, our results indicate that the prevalence of monotonic reddening and Lommel-Seeliger-type scattering for these bodies results from microscopic roughness, both in the form of structures built by the particles and roughness on the surface of the particles themselves. It follows from the singular Viking~1 observation that the surface in front of the lander was composed of semi-transparent particles, and was smooth on the scale of the particle size.

\end{abstract}


\end{frontmatter}


\section{Introduction}
\label{sec:introduction}

\begin{figure}
\centering
\includegraphics[width=7cm]{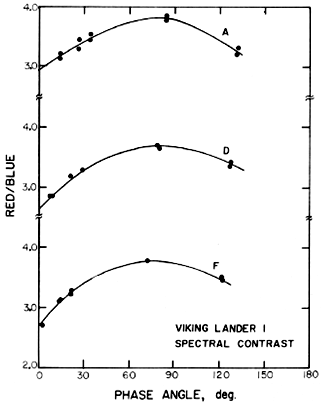}
\caption{Observations of phase reddening for Martian soil by the Viking~1 lander. ``A'', ``D'', and ``F'' refer to different patches of soil in front of the lander. Figure from \citet{G81}, \copyright1981 American Geophysical Union, reproduced with permission.}
\label{fig:Viking}
\end{figure}

The reflectance spectrum of surfaces of many solar system bodies is known to depend on the observation geometry. Often, the spectrum becomes redder (increase of the spectral slope) with increasing solar phase angle in the visible wavelength range, a phenomenon known as {\it phase reddening}. Reddening has been found in both resolved and unresolved observations of a wide variety of solar system bodies, such as the Moon \citep{G64,McC69}, asteroids \citep{T71,C02,A06,M12}, Mercury \citep{WB08}, and the moons of Uranus \citep{N87}. The opposite, phase bluing (decrease of the spectral slope), is observed more rarely \citep{R09}. A third type of spectral change that has been observed is deepening of absorption bands \citep{C02,DV07,S12,L14}. Spacecraft have always observed either phase reddening or phase bluing, with one notable exception: Viking~1 found the spectrum of the Martian soil in front of the lander to change from reddening to bluing with increasing solar phase angle \citep{G81}. This observation is so unique and relevant to this paper, that we include it here as Fig.~\ref{fig:Viking}. Meanwhile, the arched shape of the reflectance ratio is frequently encountered in laboratory experiments \citep{AF67,GV82,K10,J13,P13}.

One of the earliest laboratory experiments on phase reddening was performed by \citet{AF67}, who measured the reflectance of several silicate powders with a photo-goniometer. The authors made several fundamental observations. First, they observed how the samples brightened with decreasing particle size, a phenomenon explained by \citet{HH63} as being due to smaller particles being more translucent. Second, they found the color, expressed as the ratio of red over blue reflectance ($R/B$), to depend on albedo and thereby particle size. Third, the color of powder samples changed with phase angle, with reddening up to a certain phase angle and bluing beyond. The authors explained this in terms of the amount of scattering that takes place at the surface and the optical path length, here defined as the total distance a light ray has traveled within the particles. In a surface composed of semi-transparent particles (neither fully transparent nor fully opaque), scattering of light rays occurs at the particle interfaces, and the absorption probability depends on the path length. The absorption coefficient is a function of wavelength and, for most minerals, blue light is absorbed more strongly than red light for the same path length. At low phase angles, the average path will be relatively short for rays that escape, as most will have reflected off the surface of the topmost particles. At intermediate phase angles (around $90^\circ$), the ray trajectories are more complicated with rays penetrating deeper into medium, and the average path is longer. At large phase angles (towards $180^\circ$), most scattering events will again happen in the topmost layer so that rays will not penetrate deeply into the sample, decreasing the average path length for rays that escape. Thus one expects $R/B$ to reach a maximum at intermediate phase angles. To test this geometrical argument, numerical simulations using the geometric optics approach are appropriate. \citet{GS08} were able to reproduce a key aspect of the experiments -- reddening up to a certain phase angle and bluing beyond -- for a surface of semi-transparent particles. Their simulations show that the average path length of rays that escape the surface peaks at an intermediate phase angle, and suggest a relation with particle size.

Several fundamental questions regarding phase reddening remain unanswered, though. Are the three types of spectral change observed with increasing phase angle (reddening, bluing, and absorption band deepening) expressions of a single physical phenomenon? What can observations of phase reddening tell us about the physical properties and structure of the planetary regolith? What is the role of particle properties, like size and mineralogical composition? Why is the arch shape of the reflectance ratio so rarely observed for planetary bodies, given that it is routinely encountered in experiments? To find the answers we measured the bidirectional reflectance of samples of common rocks (basalt, granite, limestone) of different particle size, having optical properties that cover a wide range. The measurements allow us to construct phase curves for each sample at different wavelengths, and to study any color changes with phase angle. We interpret the results with help of ray-tracing simulations tuned to the experimental conditions, adopting the method of \citet{GS08}.

\section{Experiment}
\label{sec:experiment}

\subsection{Method}

The experiment was performed with the {\it Physikalisches Institut Radiometric Experiment} (PHIRE) photo-goniometer \citep{G06} at the {\it Physikalisches Institut} of the University of Bern, Switzerland, in September 2010. PHIRE has two arms that can be moved independently. One arm carries the light source, and its angle with the surface normal is the incidence angle $\iota$. The angle of the detector arm with the surface normal is the emergence, or emission, angle $\epsilon$. A particulate rock sample was distributed inside a petri dish of 8~cm diameter such that the surface was approximately smooth. The light source illuminated a spot on the sample surface with a diameter of 2.0-2.5~cm at normal incidence. The region sampled by the detector is always larger than the spot. We took care of having ample margin between the edges of the spot and the edges of the petri dish, also at large phase angles. The light source has a divergence of $1.5^\circ$, which is the typical uncertainty in phase angle of our measurements. Spot, light source, and detector were always in a plane perpendicular to the surface. The azimuth angle ($\psi$) was constant. In this plane, on one side of the surface normal the angles are defined negative, on the other side positive. For example, if the incoming angle is $\iota = 60^\circ$, then for $\epsilon = -60^\circ$ the detector arm is positioned exactly in the opposite position on the other side of the surface normal. For this geometry, the phase angle of the observation is $\alpha = |\iota - \epsilon|$.

\begin{figure}
\centering
\includegraphics[width=8cm,angle=0]{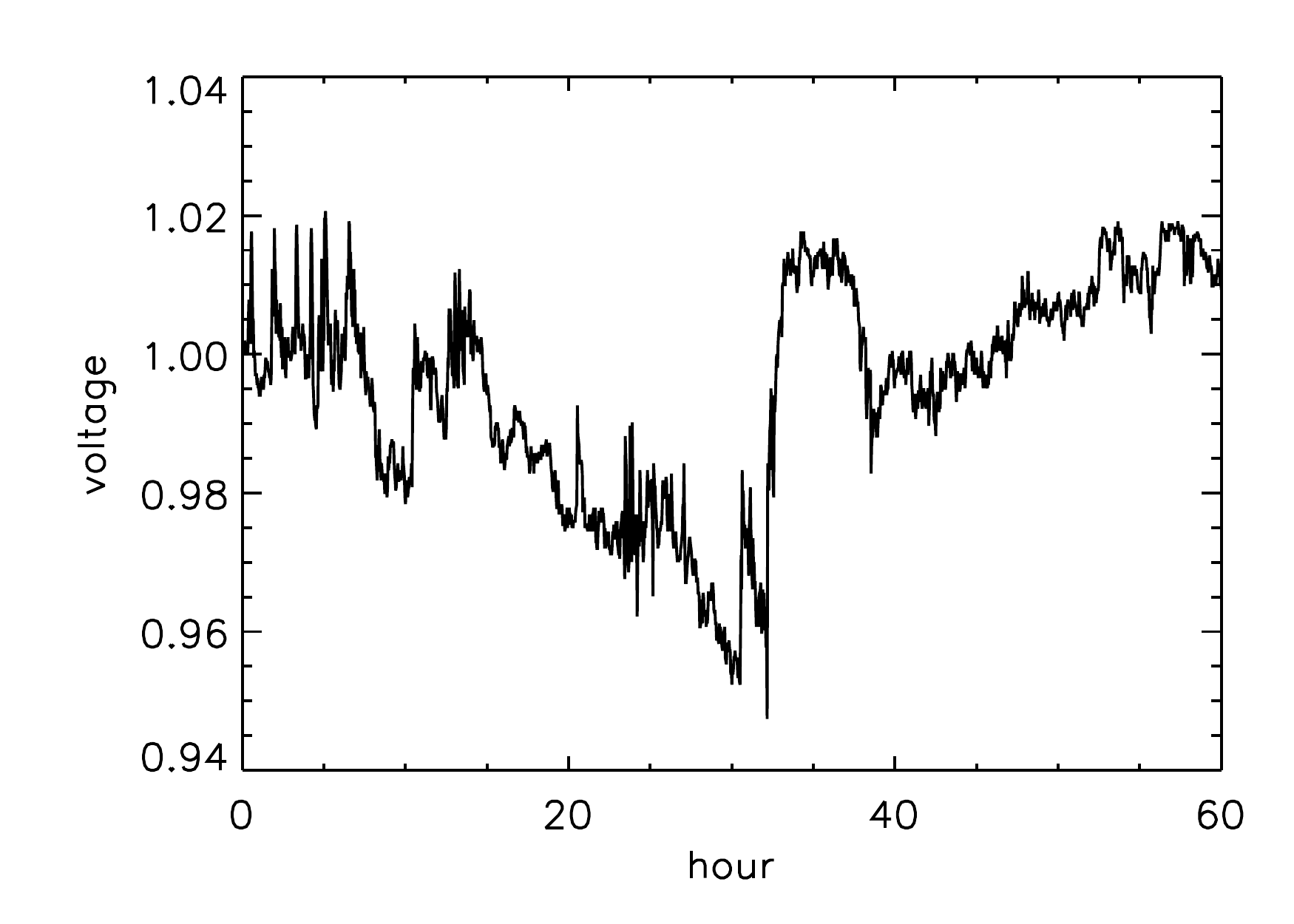}
\caption{Drift of the reflectance of the Spectralon calibration standard. The measured voltage (normalized at zero hour) is directly related to the reflectance.}
\label{fig:stability}
\end{figure}

We determine the reflectance as the radiance factor $r_{\rm F}$ (or $I/F$), which is the bidirectional reflectance of the surface relative to that of a perfectly reflecting Lambertian surface illuminated normally \citep{H81}. For the latter we adopted a square Labsphere Spectralon reference target, typically measured once or twice per day. However, after completing the experiment we determined gradual fluctuations of up to 5\% when measuring the radiance of the target almost continuously over the course of a few days (Fig.~\ref{fig:stability}), due to drift in the lamp output. A phase curve was acquired by varying the emergence angle in steps of $5^\circ$ at constant incidence angle. The error of an individual measurement was small compared to the expected drift of the lamp over the course of a phase curve acquisition (short-term drift $< 2$\%; Fig.~\ref{fig:stability}). Because of the long-term drift in lamp output we may expect differences of up to 5\% between phase curves. The reflectance spectra shown in this paper were constructed from different phase curves, and, as such, each of the data points in a spectrum has a typical uncertainty of 5\%. Measurements were generally performed at five different wavelengths: 400, 500, 600, 800, and 1000~nm using color filters with a full-width-at-half-maximum of 40~nm. As the lamp flux was comparatively low at 400~nm, achieving a reasonable signal-to-noise (S/N) could be very time consuming for dark samples. Hence, measurements for two large size classes of basalt were skipped because of time constraints. The 400~nm data for the other basalt size classes were averaged over 9 measurements to increase the S/N.

\subsection{Samples}
\label{sec:samples}

\begin{figure}
\centering
\includegraphics[width=\textwidth,angle=0]{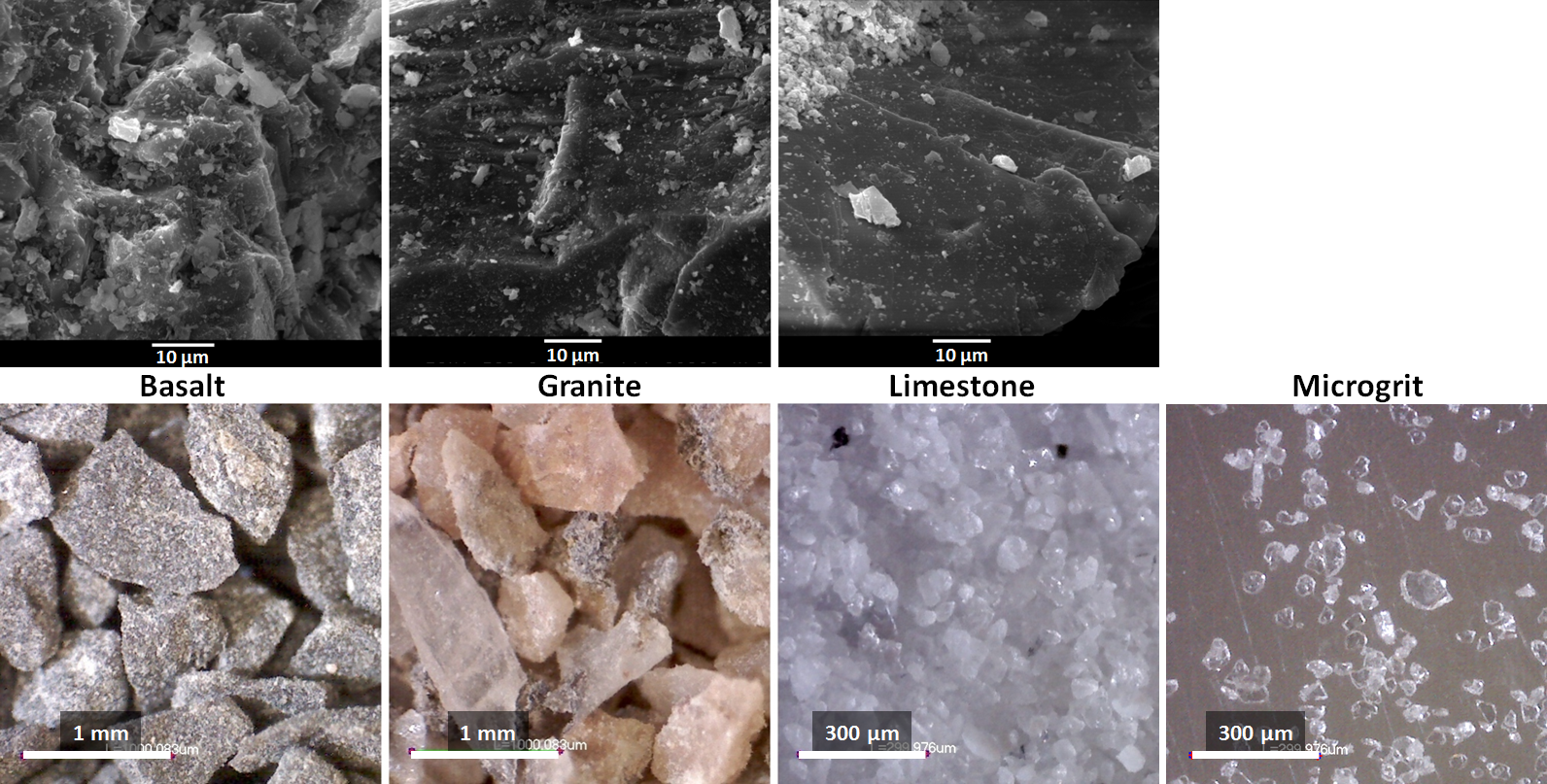}
\caption{Micrographs of the samples. {\bf Top}: SEM images. {\bf Bottom}: OM images. From left to right: basalt (500-630~\textmu m), granite (500-630~\textmu m), limestone (40-63~\textmu m), Microgrit (WCA3). We did not acquire a SEM image of Microgrit.}
\label{fig:samples}
\end{figure}

\begin{figure}
\centering
\includegraphics[width=\textwidth,angle=0]{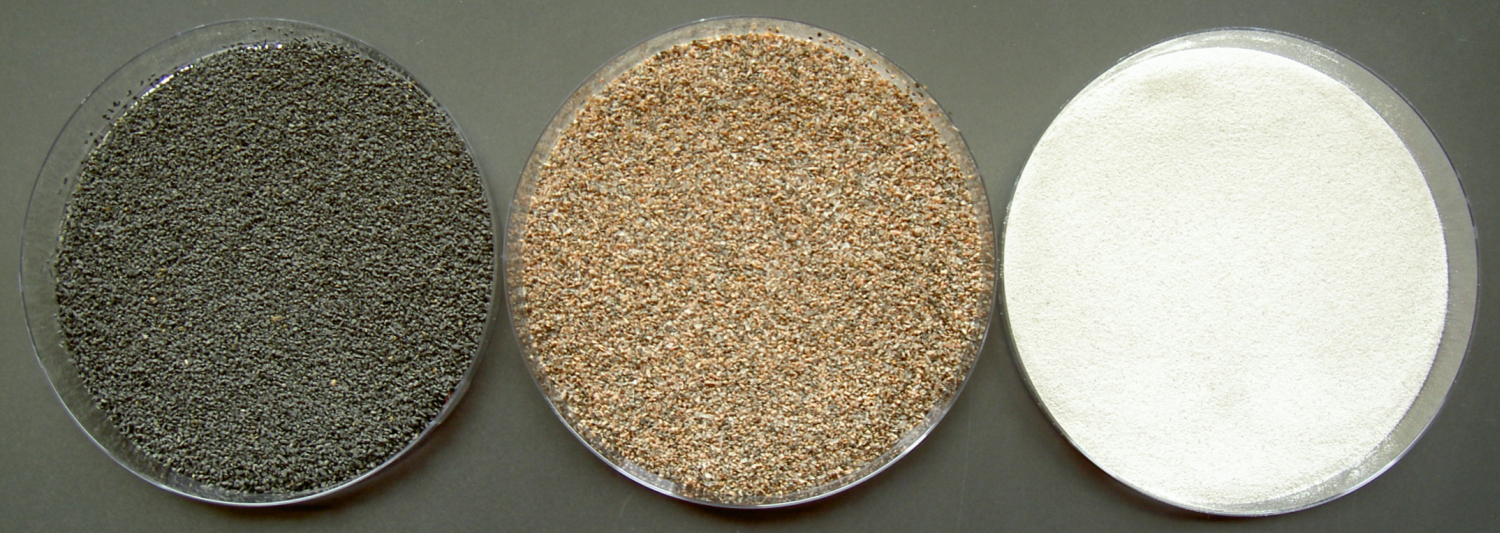}
\caption{The three rock types photographed side-by-side to show their correct relative color and brightness. From left to right: basalt (500-630~\textmu m), granite (500-630~\textmu m), and limestone (63-80~\textmu m).}
\label{fig:samples_photo}
\end{figure}

We measured the bidirectional reflectance of particulate samples of three different types of common terrestrial rocks: basalt, granite, and limestone. The samples were prepared at the DLR {\it Institute of Planetary Research} in Berlin. First, the bulk sample was crushed and ground using a jaw crusher and a ball-mill, then the ground sample was dry-sieved to several size fractions. Details of the rock types and size classes selected for our experiment are listed in Table~\ref{tab:samples}. We performed an additional characterization of the samples by imaging them with both a scanning electron microscope (SEM) and an optical microscope (OM). The micrographs in Fig.~\ref{fig:samples} show the particles at sub-millimeter scale. Because the OM images do not show the rock samples at their correct relative brightness and color, we photographed them side-by-side in petri-dishes (Fig.~\ref{fig:samples_photo}). For reference, we also performed measurements on commercially available Microgrit (Al$_2$O$_3$). Microgrit powder is pure white, with disc-like and highly transparent particles. We measured the WCA3 variety, with a typical disc diameter of 20-40~\textmu m as judged from the OM image (Fig.~\ref{fig:samples}). As we did not include it in Fig.~\ref{fig:samples_photo}, we report that Microgrit appeared a little whiter and brighter than limestone.

\begin{figure}
\centering
\includegraphics[width=\textwidth,angle=0]{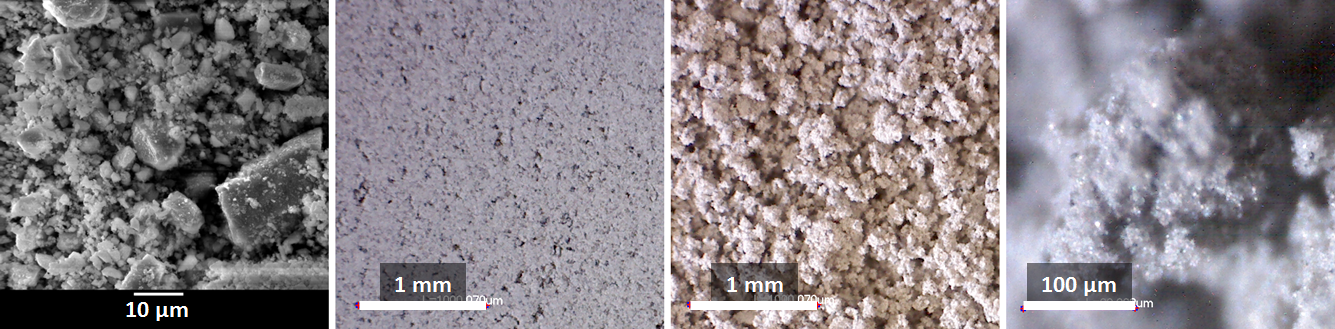}
\caption{Micrographs of the powder basalt samples. From left to right: SEM image, OM image of the pressed sample, OM image of the sprinkled sample, OM image of the sprinkled sample.}
\label{fig:basalt_powder}
\end{figure}

Four size classes of basalt were selected, the smallest of which can be considered a powder. The coarse basalt particles were very dark and opaque. Their surface was very rough; the SEM images show abundant structure on the micrometer scale (Fig.~\ref{fig:samples}). The basalt powder was extremely fine; the SEM images show particle sizes from around 10~\textmu m to much smaller than 1~\textmu m. The powder sample was prepared in two different ways: sprinkled and pressed (Fig.~\ref{fig:basalt_powder}). First, it was sprinkled through a plastic sieve on the inside of a petri dish cover. After measurements were performed this sample was pressed by hand using the petri dish bottom, and the same set of measurements was repeated. While the pressed sample was smooth on the scale of the particle size, the sprinkled particles spontaneously built little tower structures. These structures were first reported by \citet{HH63}, and dubbed {\it fairy castles}. A close-up of one such fairy castle in Fig.~\ref{fig:basalt_powder} shows its large size compared that of the particles of which it is composed. Thus, the sprinkled basalt powder sample was rough on a microscopic scale, to which we refer as {\it microscopic roughness}. In the planetary photometry literature, different definitions of {\it macroscopic roughness} can be found. \citet{FV83} define macroscopic roughness as ``roughness not reproducible in the laboratory measurements, such as that due to the presence of pebbles, rocks, boulders, craters, and ridges''. The \citet{H81} macroscopic roughness is on a scale that is ``small compared with the size of the area being examined (typically of the order of several hundred square kilometers) but large compared to the extinction mean free path in the medium''. Both definitions suffer from the fact that the scale is not very well defined. The former depends on the size of one's laboratory, whereas the latter is estimated to be from 1~mm to 1~m, based on the argument that the regolith must be cohesive due to the small average particle size. In this paper, we therefore adhere to a more ``classical'' definition of macroscopic roughness as the opposite of microscopic roughness, which is roughness that is indistinct without the use of a microscope.

The basalt powder was bright relative to the coarse basalt particles due to the small particle size. The reddish granite samples showed a large diversity in particle type; some were transparent, some opaque. The surface of the opaque particles was rough on a micrometer scale, like that of the coarse basalt particles, while that of the transparent particles was relatively smooth. The SEM image (Fig.~\ref{fig:samples}) shows an example of the latter. The very bright limestone samples were mainly composed of transparent particles interspersed with a few very dark particles. The surface of the transparent particles was very smooth, as evident in the SEM image. In the top left corner a very rough patch is visible, which is probably an example of the very dark material that was attached to some of the transparent particles, as seen in the optical microscope images. The limestone particles were electrostatic, especially in the smallest size class. This was seen for neither basalt nor granite. While this would lead to a lumpy surface texture, limestone particles did not build fairy castle structures. The coarse basalt and granite particles were relatively large, so that is was possible to create smooth surfaces on the scale of the particle size by gently shaking the petri dish. The limestone samples were difficult to smoothen due to the small particle size. We tried two different methods: the petri-dishes of the 10-32 and 63-80~\textmu m samples were shaken, whereas the 40-63~\textmu m sample was sprinkled through a plastic sieve. The sprinkled sample was less lumpy than the shaken ones, but not necessarily smoother.

\subsection{Results}
\label{sec:results}

\subsubsection{Phase Curves}

\begin{figure}
\centering
\includegraphics[width=8cm,angle=0]{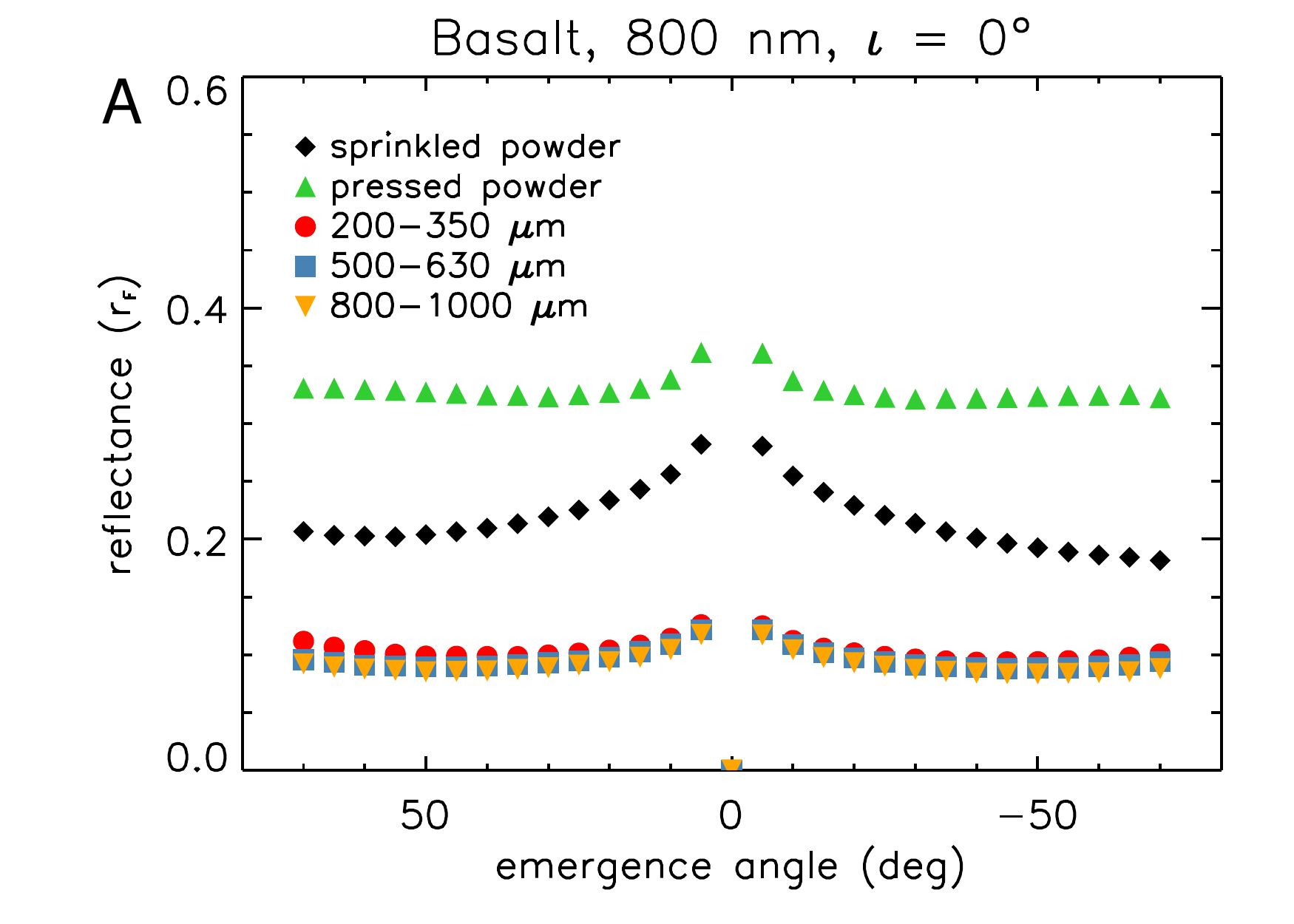}
\includegraphics[width=8cm,angle=0]{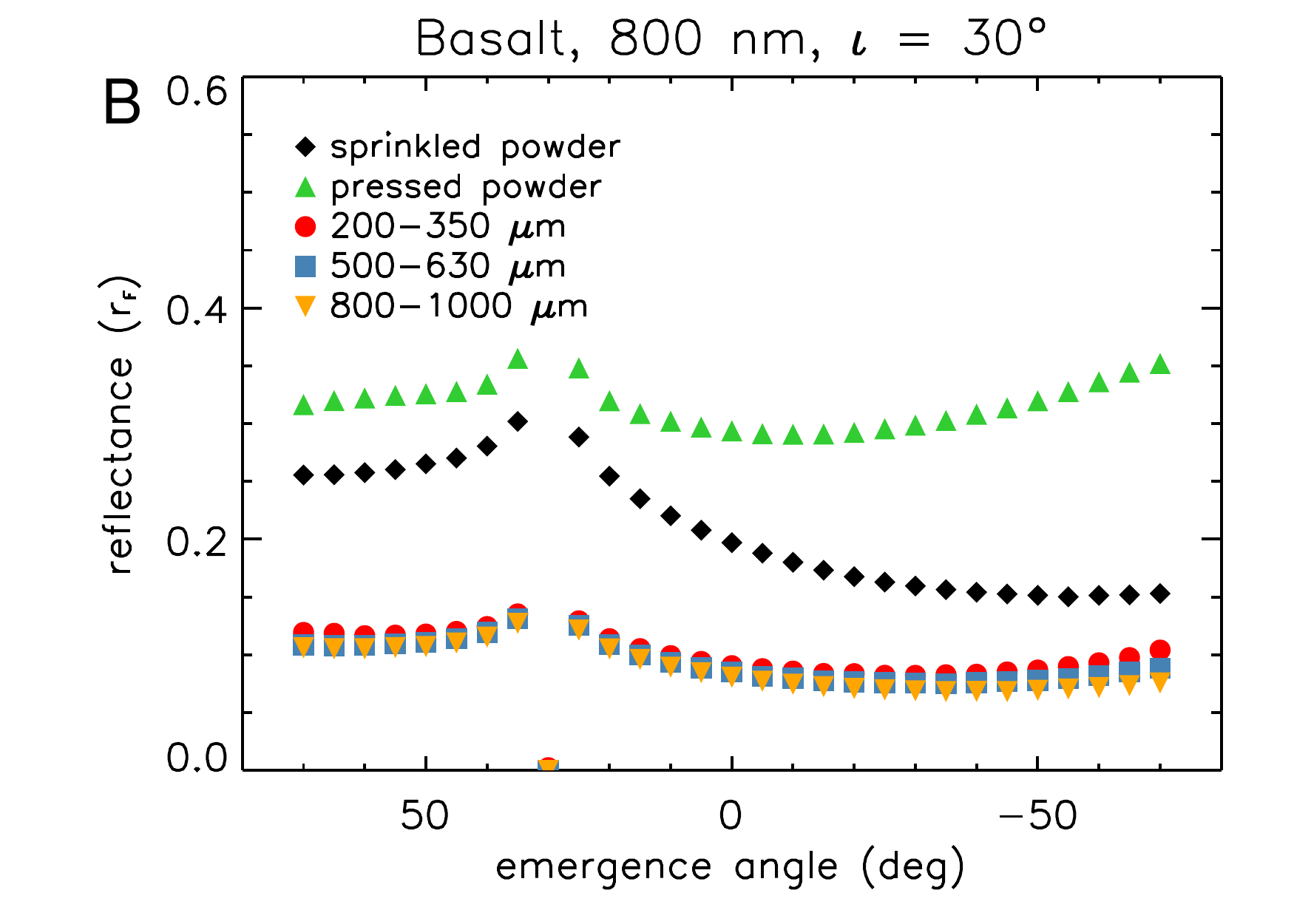}
\includegraphics[width=8cm,angle=0]{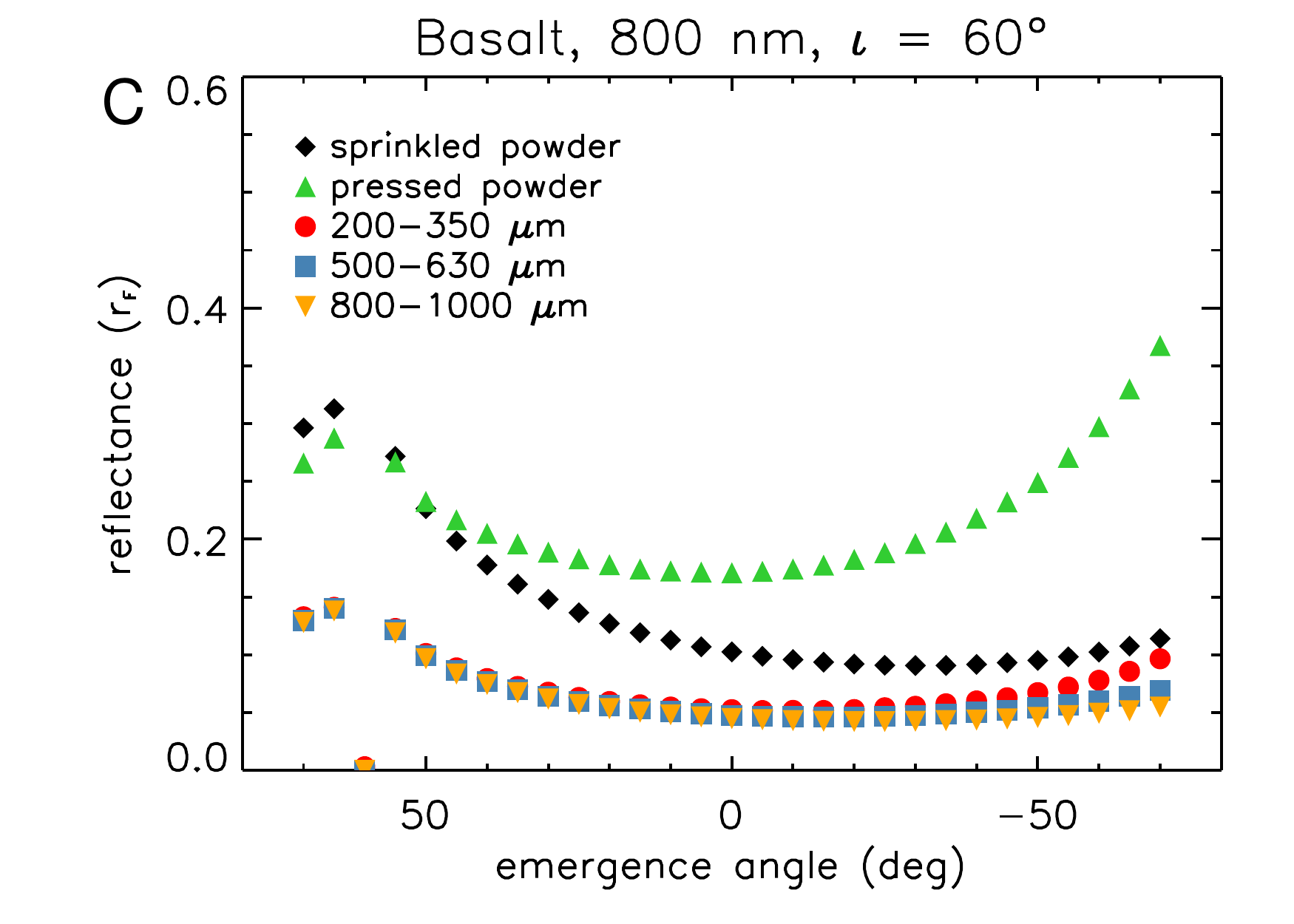} \\
\includegraphics[width=8cm,angle=0]{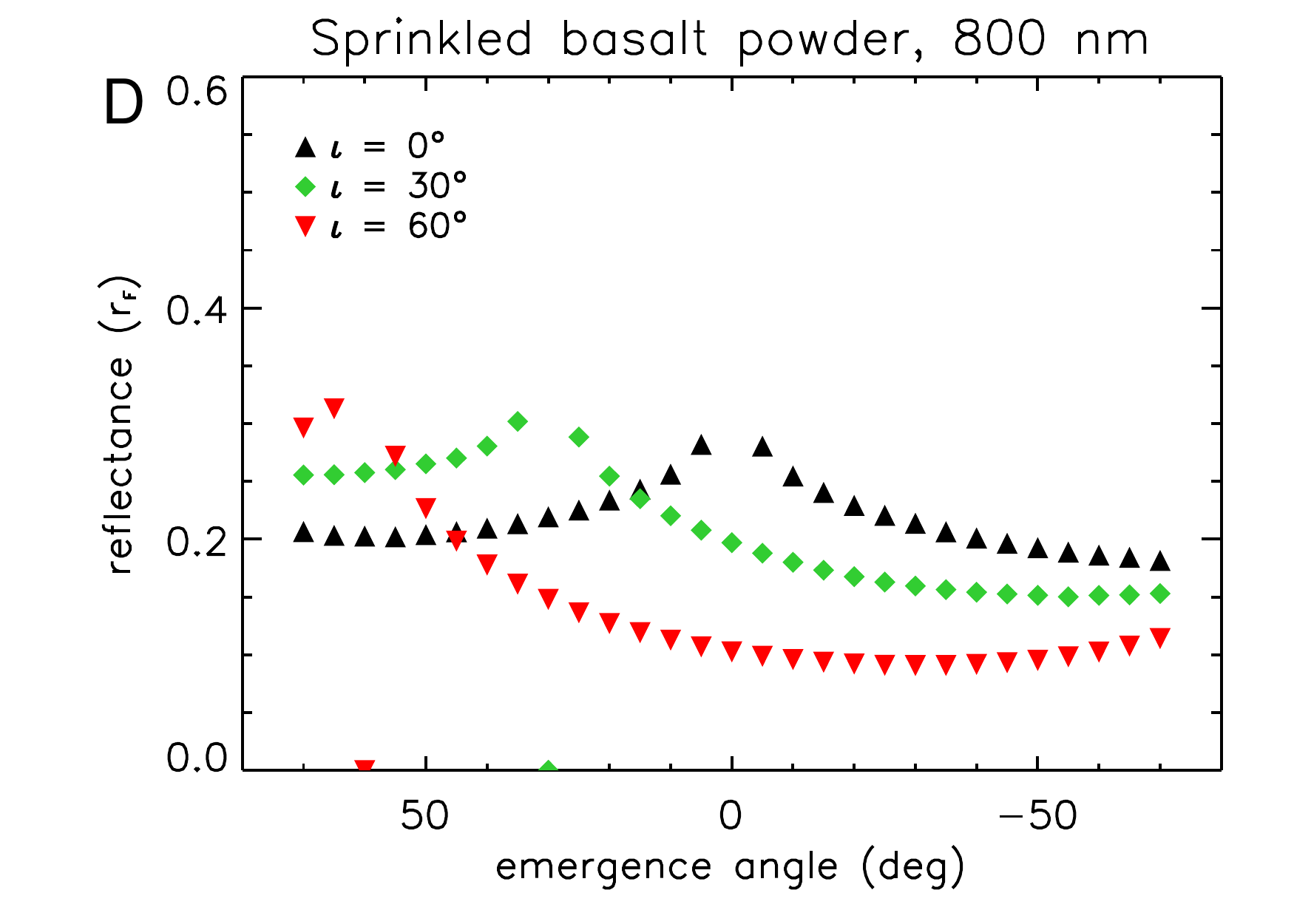}
\includegraphics[width=8cm,angle=0]{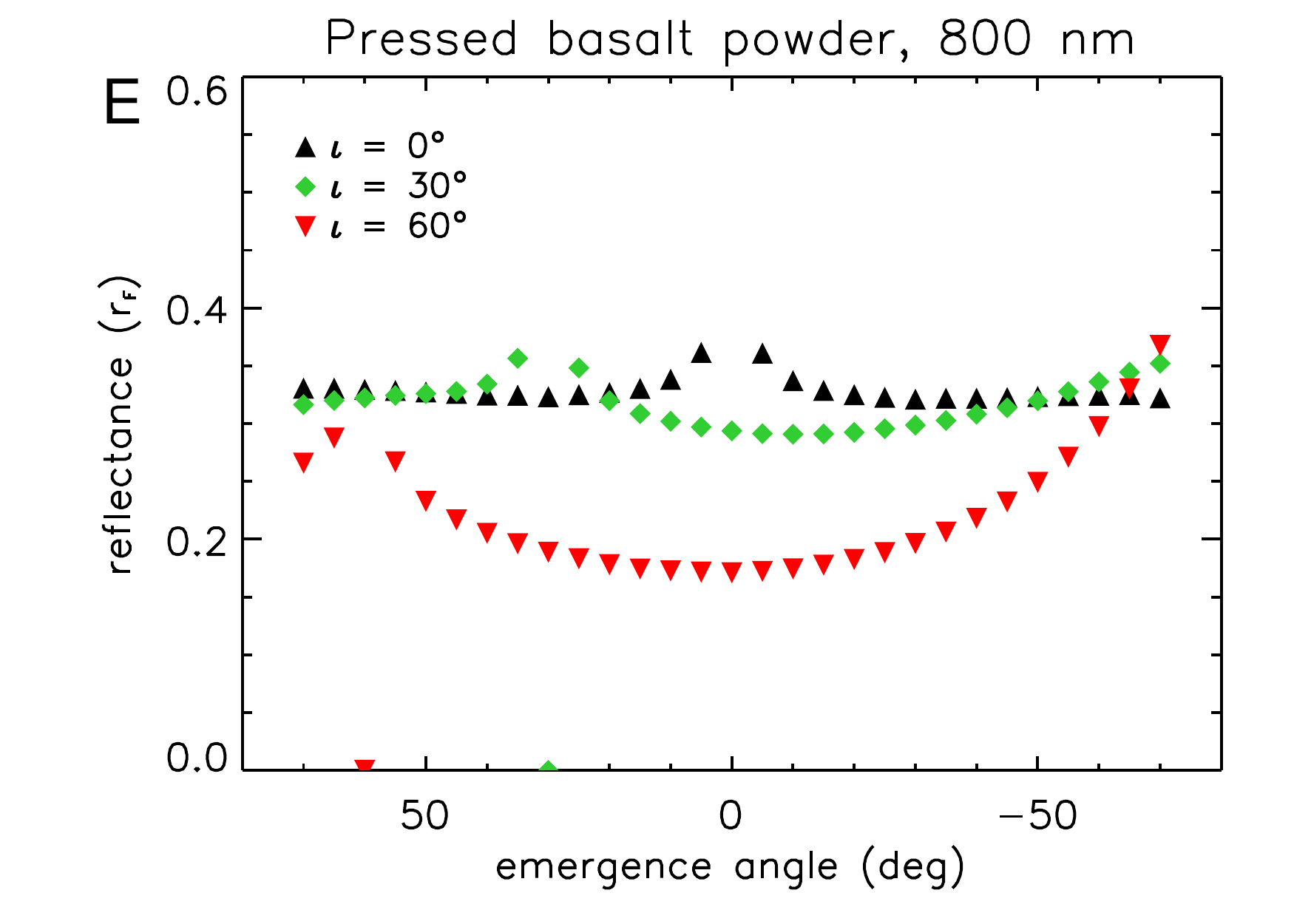}
\caption{Phase curves associated with the basalt samples at 800~nm. {\bf A}:~all sizes, $\iota = 0^\circ$, {\bf B}:~all sizes, $\iota = 30^\circ$, {\bf C}:~all sizes, $\iota = 60^\circ$, {\bf D}:~sprinkled powder, {\bf E}:~pressed powder.}
\label{fig:basalt_phase_curves}
\end{figure}

Reflectance phase curves of the basalt, granite, and limestone samples were obtained at three different incidence angles: $\iota = 0^\circ$, $30^\circ$, and $60^\circ$. If the sample surface is flat and horizontal, the $\iota = 0^\circ$ curves will be symmetric. The measurements were performed at five different wavelengths: 400, 500, 600, 800, and 1000~nm, but 400~nm phase curves were only acquired for selected samples due to time constraints. Here we show only the phase curves for 800~nm; those at the other wavelengths are comparable (the dependence of the phase curve on wavelength is described in detail in Sec.~\ref{sec:phase_reddening}). The results for basalt are shown in Fig.~\ref{fig:basalt_phase_curves}. The $\iota = 0^\circ$ profiles (Fig.~\ref{fig:basalt_phase_curves}A) are indeed symmetric, albeit not perfectly, indicating that the surfaces were fairly level. The powder reflectance is much higher than that of the bigger particles, demonstrating the well-known increase of reflectance with decreasing particle size \citep{HH63,AF67}. Smaller particles, being more transparent, are also more forward scattering. This is seen especially well in the $\iota = 60^\circ$ phase curve of pressed basalt powder (Fig.~\ref{fig:basalt_phase_curves}C), where the reflectance increases strongly towards more negative emergence angles (larger phase angles). Forward scattering is strongly reduced for the sprinkled powder through the presence of the ``fairy castle'' structures described in Sec.~\ref{sec:samples}. The repression can be understood in terms of the higher opacity of these structures, leading to mutual shadowing. Real-world examples of how pressing a powder surface increases forward scattering are the Apollo Lunar rover tracks \citep{K11} and the Spirit and Opportunity Mars rover tracks \citep{J06a,J06b}.

\begin{figure}
\centering
\includegraphics[width=8cm,angle=0]{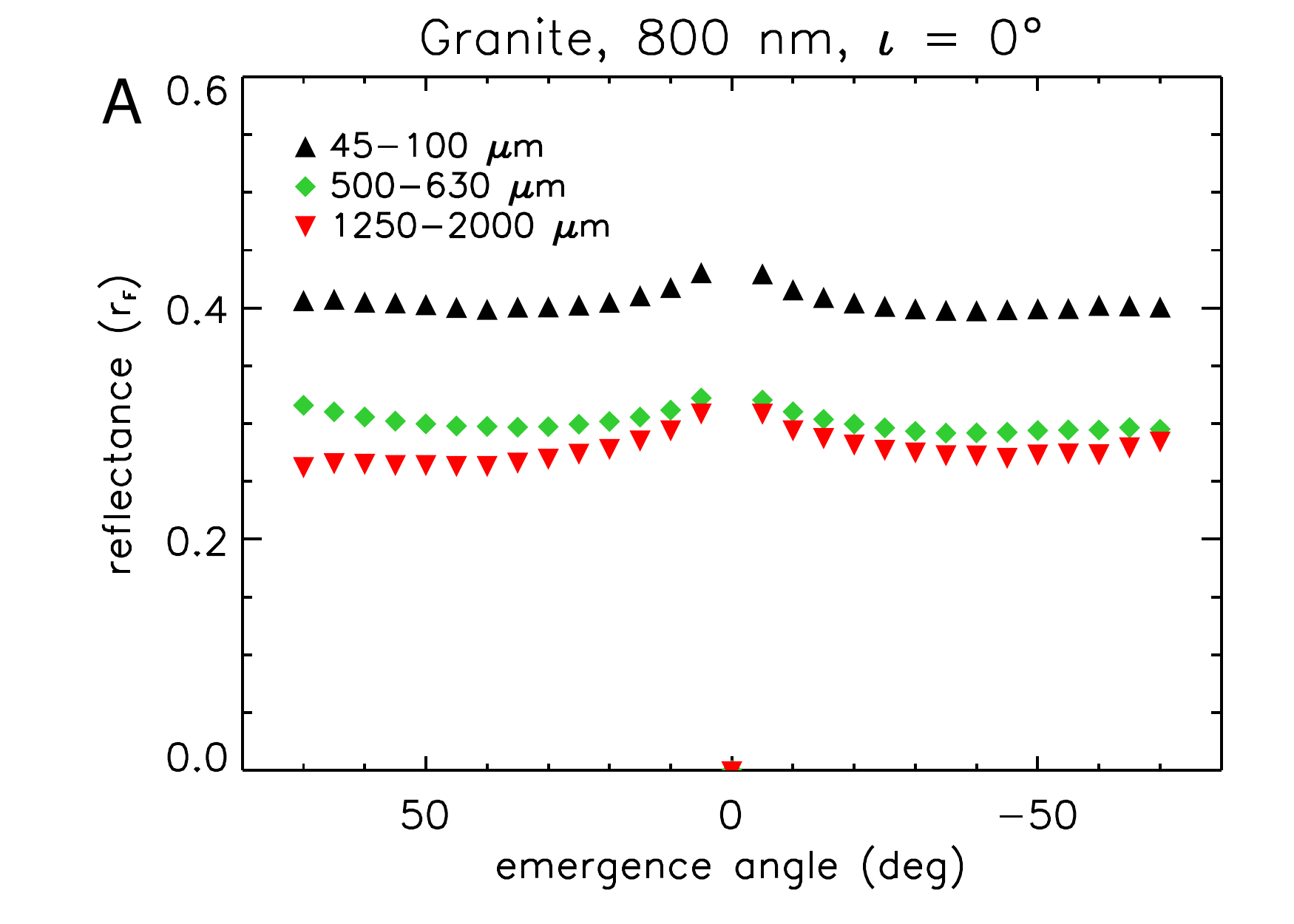}
\includegraphics[width=8cm,angle=0]{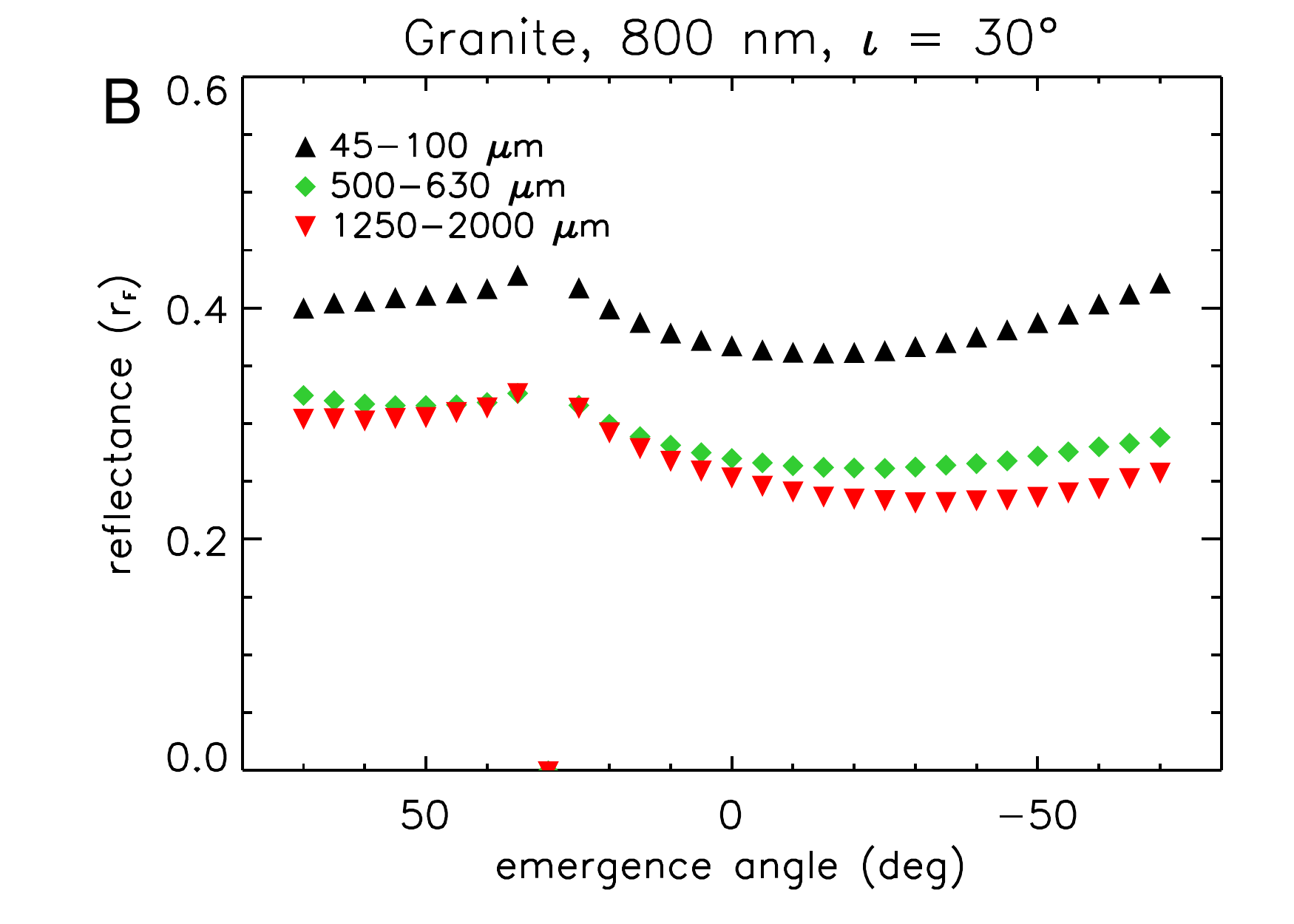}
\includegraphics[width=8cm,angle=0]{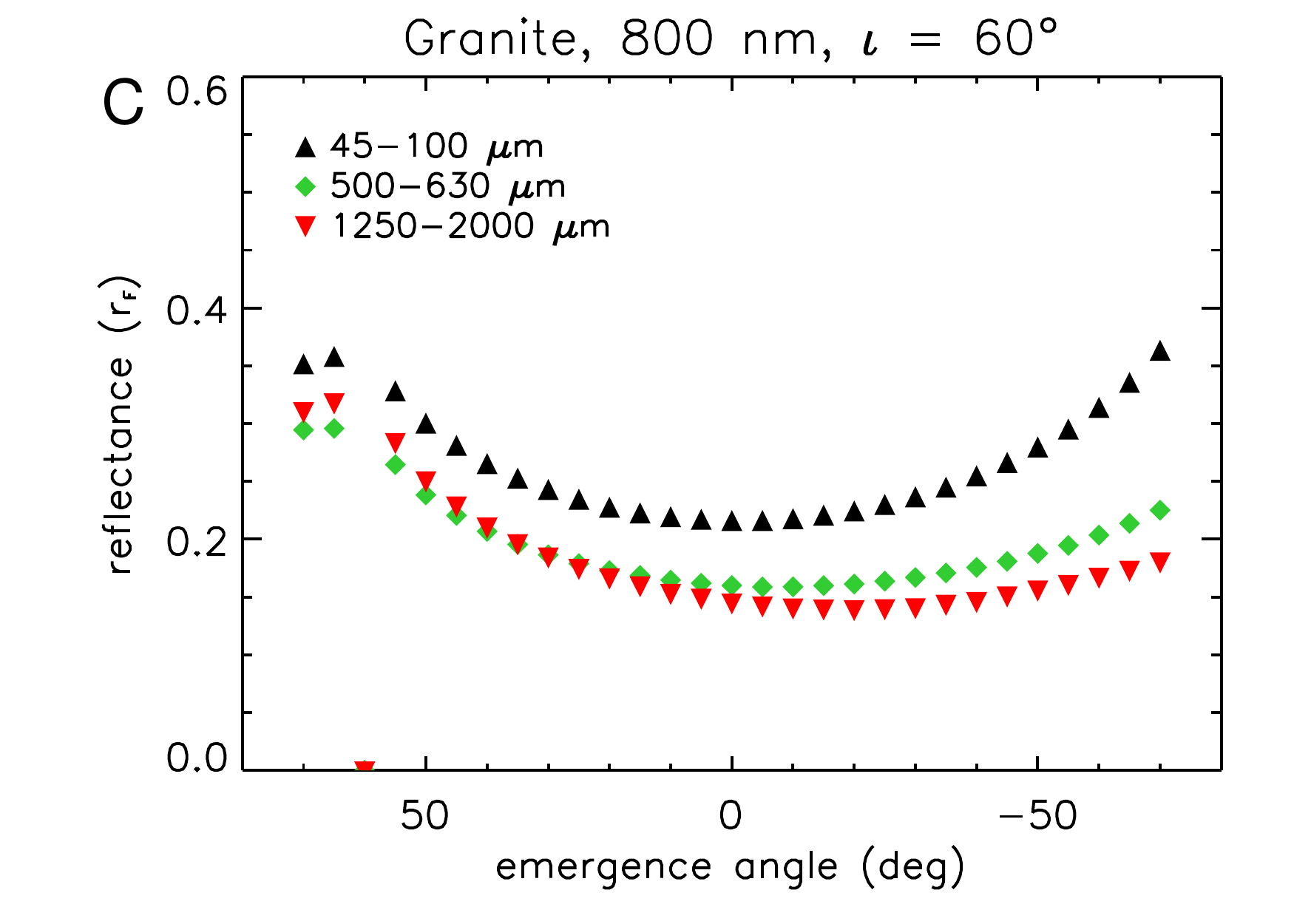}
\caption{Phase curves associated with the granite samples of all sizes at 800~nm. {\bf A}:~$\iota = 0^\circ$, {\bf B}:~$\iota = 30^\circ$, {\bf C}:~$\iota = 60^\circ$.}
\label{fig:granite_phase_curves}
\end{figure}

The phase curves of the granite samples are shown in Fig.~\ref{fig:granite_phase_curves}. The $\iota = 0^\circ$ profiles (Fig.~\ref{fig:granite_phase_curves}A) show some asymmetries, indicating minor non-flatness. The $\iota = 60^\circ$ profiles (Fig.~\ref{fig:granite_phase_curves}C) are strongly curved and resemble those of the pressed basalt powder sample, indicating that, on average, the opacity of the granite particles is comparable to that of the much smaller basalt powder particles. This upward turn of the reflectance for larger phase angles is stronger for smaller particle sizes. The overall reflectance of the smallest particle size sample (45-100 \textmu m) is significantly higher than that of the two larger size classes.

\begin{figure}
\centering
\includegraphics[width=8cm,angle=0]{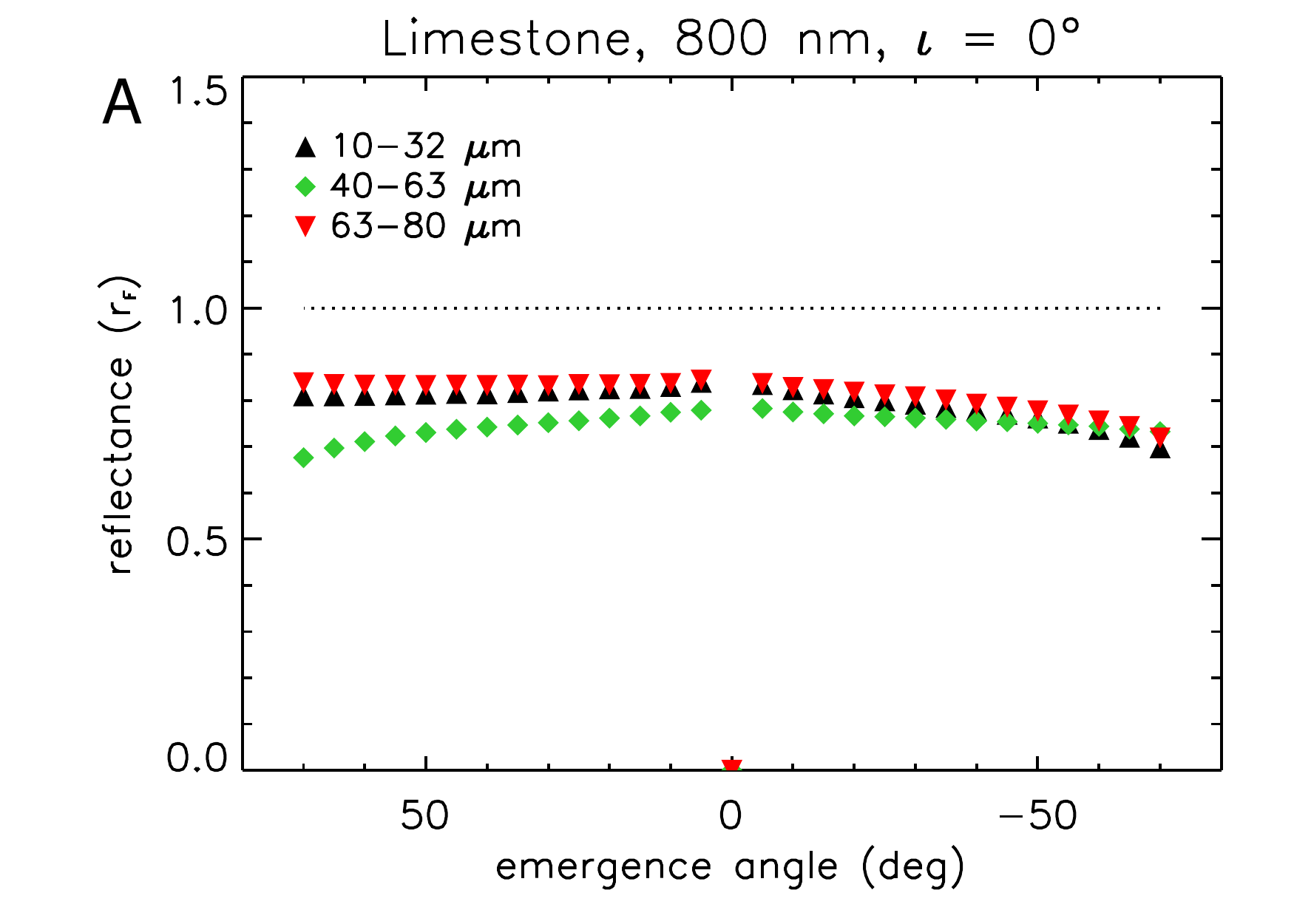}
\includegraphics[width=8cm,angle=0]{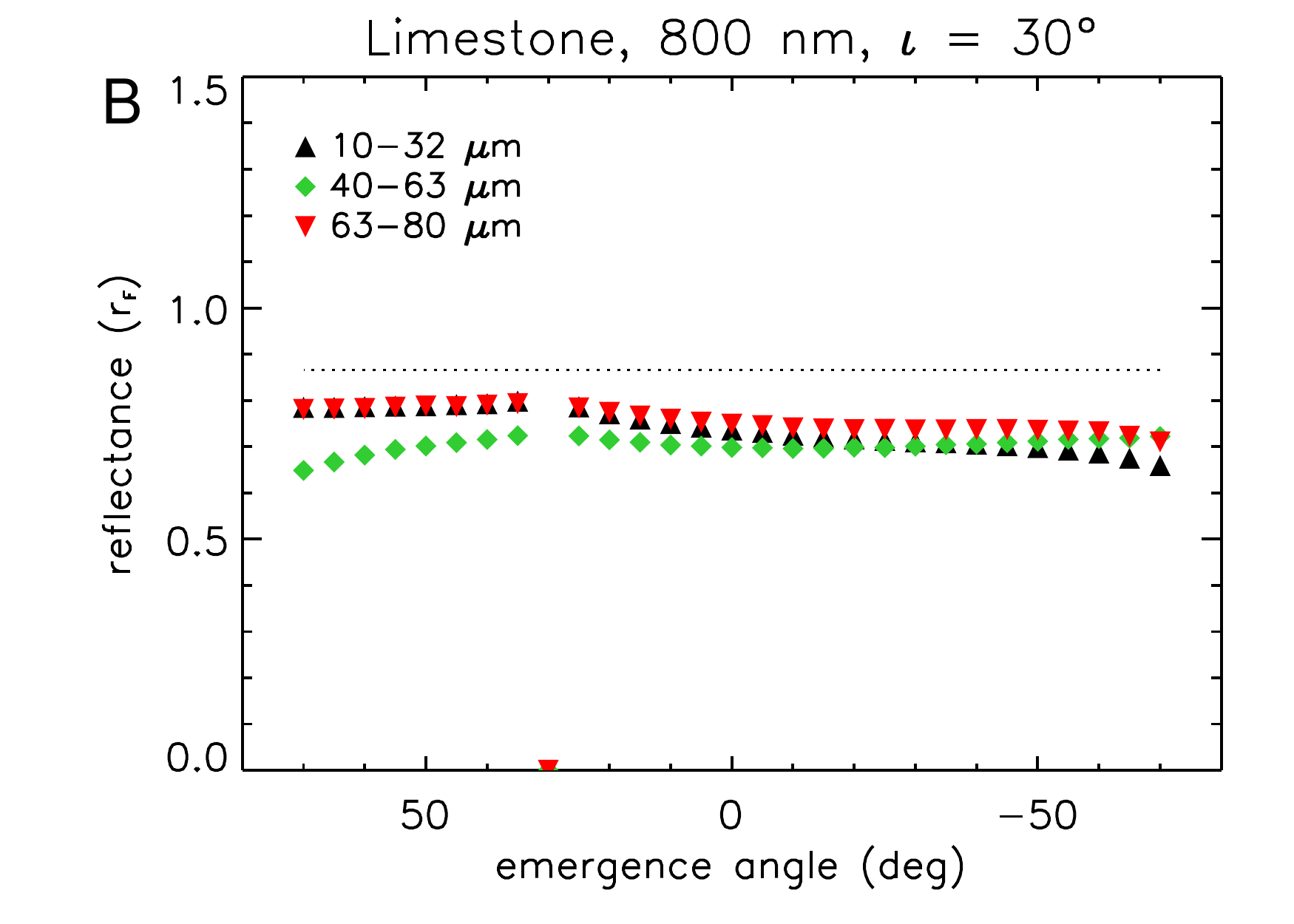}
\includegraphics[width=8cm,angle=0]{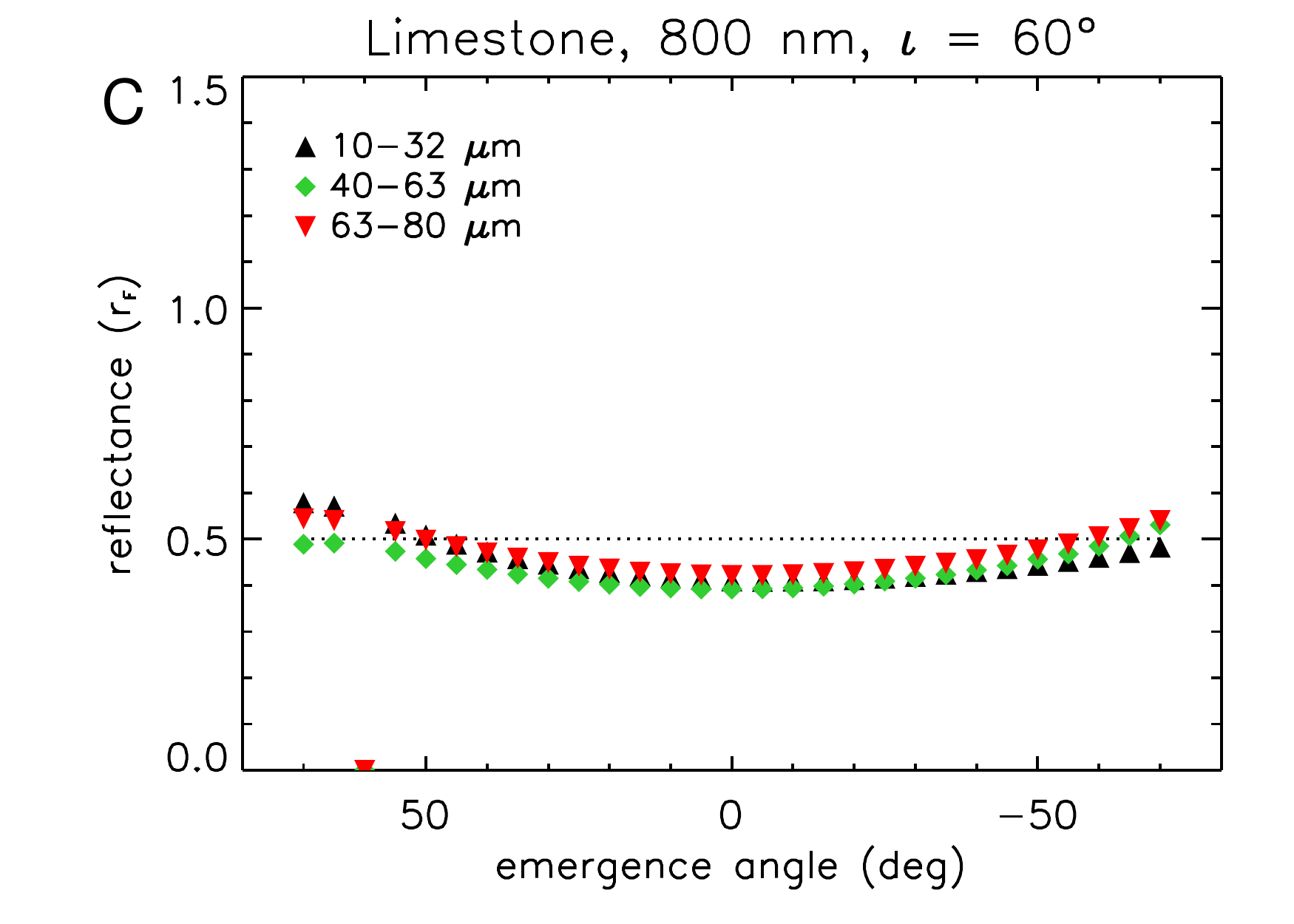}
\caption{Phase curves associated with the limestone samples of all sizes at 800~nm. {\bf A}:~$\iota = 0^\circ$, {\bf B}:~$\iota = 30^\circ$, {\bf C}:~$\iota = 60^\circ$. The dotted line denotes the reflectance of a Lambertian surface.}
\label{fig:lime_stone_phase_curves}
\end{figure}

The limestone phase curves are shown in Fig.~\ref{fig:lime_stone_phase_curves}. The samples were difficult to flatten, which shows in the asymmetries in the $\iota = 0^\circ$ profiles (Fig.~\ref{fig:lime_stone_phase_curves}A). The three particle size classes have very similar phase curves, which are generally rather flat. The $\iota = 60^\circ$ profiles (Fig.~\ref{fig:lime_stone_phase_curves}C) are weakly curved, and much shallower than the granite profiles. The sample with the smallest grain size appears to be less forward scattering than the others, contrary to what we observe for basalt and granite. However, this effect is most likely not real. The differences between the size classes are very small, and the unevenness of the surface and the drift in lamp output (Fig.~\ref{fig:stability}) can easily be responsible. The figures also include reflectance profiles expected for a perfect Lambertian surface. We find that the limestone samples reflect close to Lambertian, especially at higher incidence angles.

Comparing the phase curves of the different materials and grain sizes, we find that only those samples are forward scattering that have a surface of semitransparent particles (neither fully transparent nor fully opaque) that is smooth on the scale of the particles. Examples of these are the surfaces composed of pressed basalt powder and 45-100~\textmu m granite particles.

\subsubsection{Constant phase}

In a special measurement series, we measured the reflectance of all sample materials at the constant, low, phase angle of $\alpha = 10^\circ$, while varying the incidence and emergence angles. In this way we can study what is often called the ``limb darkening'' behavior of each material \citep{H81}. This term comes from the fact that at low phase angles, high albedo planetary bodies, like the icy satellites of Saturn, exhibit a distinct darkening towards the limb \citep{B84}. However, silicate bodies like the Moon do not \citep{F60}. The reflectance profile over the planetary disk is also called the {\it disk function}, which describes how the reflectance depends on incidence and emergence angle for a certain phase angle. One can {\it photometrically correct} an image of a planetary surface by dividing by the disk function, which removes brightness variations due to topography. To analyze our data we use a simple form of the \citet{H81} model. We only use the core radiative transfer model for well-dispersed isotropic scatterers. We omit the shadow hiding or coherent backscatter terms because we are both far enough from phase angle zero for coherent backscatter to be (presumably) negligible and close enough to zero for shadows to be relatively unimportant. We also do not include the macroscopic roughness term introduced by \citet{H84} to keep it simple. The model reflectance (radiance factor) is
\begin{equation}
r_{\rm F} = \frac{w}{4}\frac{\mu_0}{\mu_0 + \mu} H(w,\mu_0) H(w,\mu),
\label{eq:model_radfac}
\end{equation}
in which $w$ is the single scattering albedo, $\mu_0 = \cos \iota$, and $\mu = \cos \epsilon$. The $H$-function is
\begin{equation}
H(w,\mu) = \frac{1 + 2 \mu}{1 + 2 \mu \sqrt{1-w}}.
\end{equation}
The single scattering contribution to the reflectance is
\begin{equation}
(r_{\rm F})_{\rm S} = \frac{w}{4}\frac{\mu_0}{\mu_0 + \mu}.
\end{equation}
For low $w$ the reflectance converges to the Lommel-Seeliger law ($r_{\rm F} \sim \mu_0 / (\mu_0 + \mu)$). For high $w$ the multiple scattering term $H(w,\mu_0) H(w,\mu)$ dominates, and the reflectance is more Lambertian.

\begin{figure}
\centering
\includegraphics[width=8cm,angle=0]{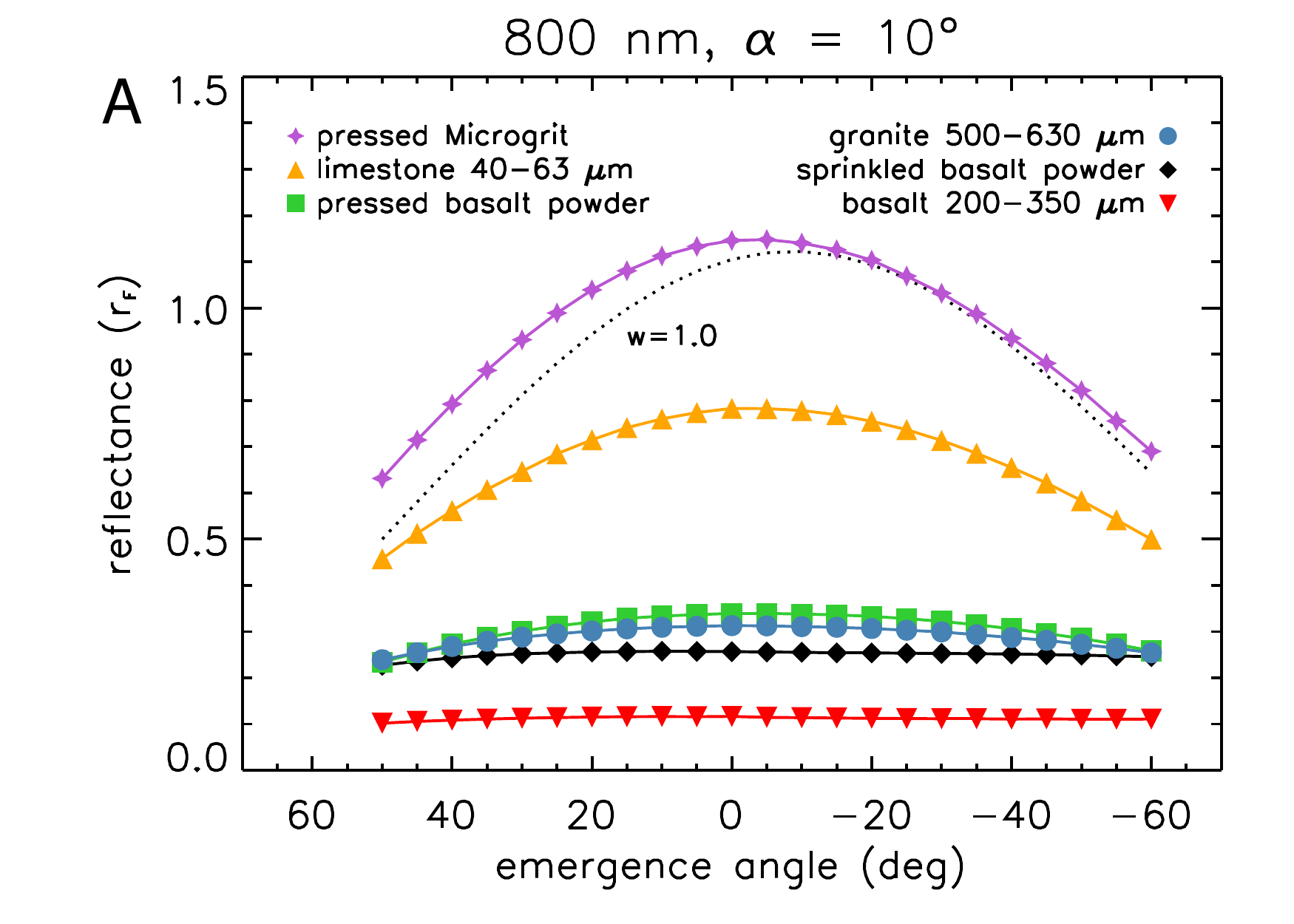}
\includegraphics[width=8cm,angle=0]{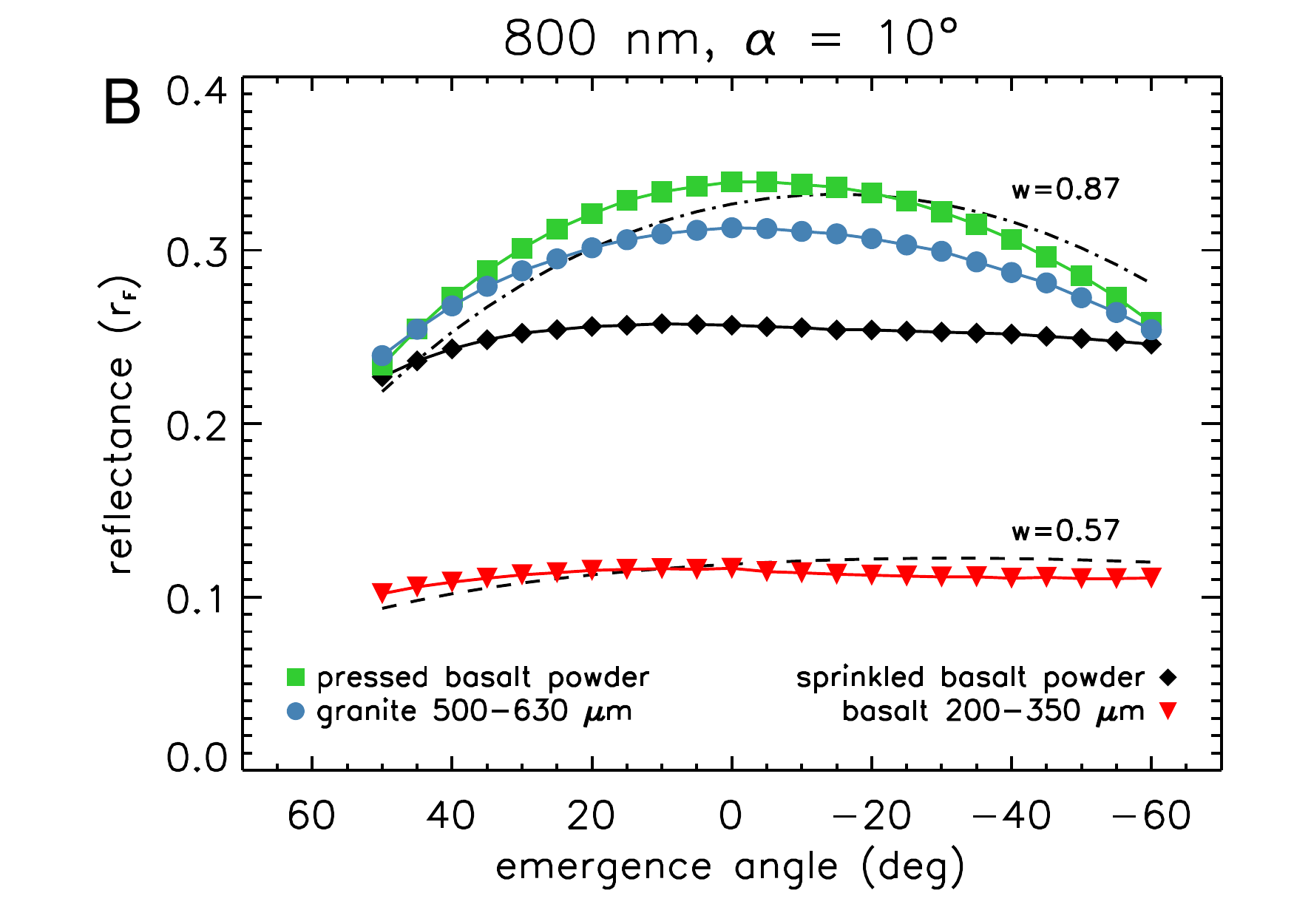}
\caption{Reflectance profiles of all materials compared for a constant phase angle of $\alpha = 10^\circ$ at 800~nm ($\iota = \epsilon + 10^\circ$). The black lines are model profiles (Eq.~\ref{eq:model_radfac}) for different values of the single scattering albedo $w$ (indicated). {\bf B} is a close up of the lower part of {\bf A}.}
\label{fig:constant_phase}
\end{figure}

In Fig.~\ref{fig:constant_phase} we compare measurements of one size class per mineral, including the basalt powder in pressed and sprinkled form. At this phase angle ($\alpha = 10^\circ$), the coarse basalt sample is perceived to be the darkest, with the granite sample being almost three times as bright (compare Fig.~\ref{fig:samples_photo}). The pressed powdered basalt sample is slightly brighter than the granite, demonstrating the well-known effect of increasing reflectance for decreasing size of semi-opaque particles. Much brighter is the limestone sample, and the Microgrit is brightest of all. The reflectance profiles for constant phase angle essentially come in two different shapes: ``curved'' for limestone, Microgrit, granite, and the pressed basalt powder, and ``flat'' for coarse basalt and the sprinkled basalt powder. The figure also includes model profiles, calculated for different values of the single scattering albedo $w$ in Eq.~\ref{eq:model_radfac}. The reflectance of the brightest sample, Microgrit, is modeled well with $w=1.0$. As the Microgrit particles are almost fully transparent, the scatterers in this surface are not the particles themselves but the particle-air and particle-particle interfaces. A perfectly spherical hypothetical planet covered by a smooth layer of Microgrit would exhibit strong limb darkening at low phase angle. The reflectance of the darkest sample, coarse basalt, is modeled well with $w=0.57$. Its reflectance profile is relatively flat, and a hypothetical planet covered by coarse basalt would exhibit little limb darkening at low phase angle. The model single scattering albedo of the darkest sample is relatively high, and single and multiple scattering contribute about equally to the total reflectance. However, the single scattering albedo in Eq.~\ref{eq:model_radfac} is overestimated because, being opaque, the particles are strongly backscattering. The powder basalt consists of particles so small that they are semi-transparent. The pressed powder reflectance is modeled reasonably well with $w=0.87$, a value in between those of Microgrit and coarse basalt. However, the sprinkled basalt reflectance is fundamentally different. Its profile is as flat as that of coarse basalt, and cannot be modeled with Eq.~\ref{eq:model_radfac}. This is due to the presence of tiny structures on the surface that form spontaneously when sprinkling, i.e.\ the \citet{HH63} fairy castles. The flattening of the profile (disk function) by this kind of microscopic roughness can be understood in the following way. The sprinkled surface appears as if it is covered by miniature hills, each presenting a surface with many different combinations of incidence and emission angles. While a hypothetical planet with a surface of pressed basalt powder will show considerable limb darkening because of the increasing incidence and emission angles towards the limb, roughness homogenizes these angles across the disk, and thereby the reflectance \citep{H81}.

\begin{figure}
\centering
\includegraphics[width=8cm,angle=0]{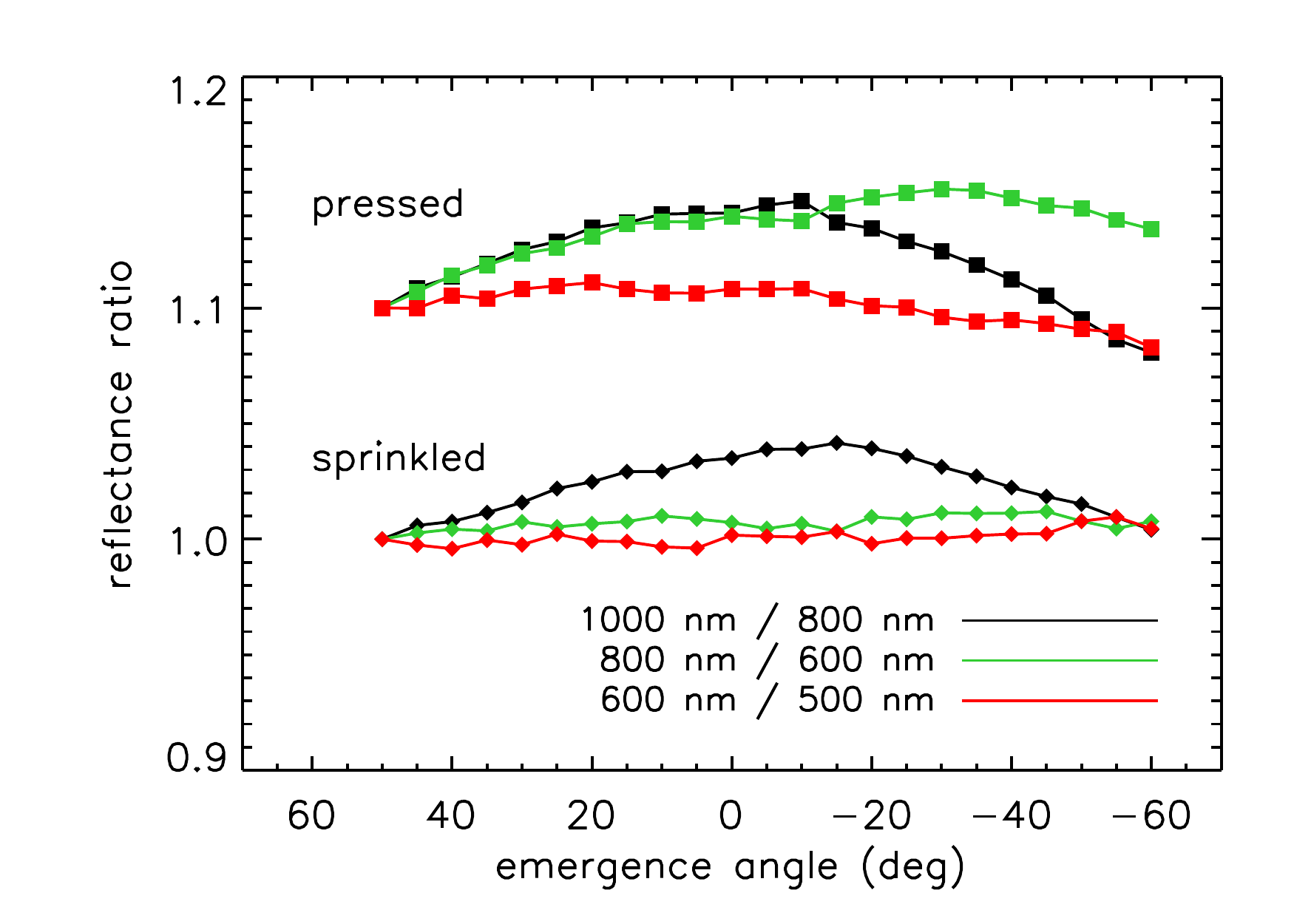}
\caption{Ratio profiles for the pressed and sprinkled basalt powder reflectances measured at a constant phase angle of $\alpha = 10^\circ$ ($\iota = \epsilon + 10^\circ$). The ratios are calculated for the three different wavelength combinations indicated in the legend. The profiles are normalized at $\epsilon = 50^\circ$, with the pressed basalt profiles offset by 0.1 for clarity}
\label{fig:color_gradient}
\end{figure}

Next, we investigate how the fairy castles affects the wavelength dependence of limb darkening. Figure~\ref{fig:color_gradient} shows systematic color changes for pressed basalt with changing emergence angle, as the reflectance ratio profiles are arch-shaped for the 1000~nm / 800~nm and 800~nm / 600~nm ratios, and possibly the 600~nm / 500~nm ratio. This means that a hypothetical planet covered by pressed basalt would exhibit a subtle color gradient from center to limb when observed at low phase angle, the limb being slightly bluer. This is as expected. As we just demonstrated, the disk function of a smooth surface is more Lommel-Seeliger-like for low albedo, whereas for high albedo it is more Lambert-like. For basalt powder, the albedo increases with wavelength up to 800~nm (Fig.~\ref{fig:basalt_phase_reddening}A), and therefore the ratio of the profiles at different wavelengths is not constant. In the presence of fairy castles, however, this ratio is constant for the 800~nm / 600~nm and 600~nm / 500~nm ratios in Fig.~\ref{fig:color_gradient}. Thus, a hypothetical planet covered by sprinkled basalt would not have a visible color gradient from center to limb. This can be understood in the same way as the flattening of the disk function, i.e.\ in terms of homogenization of the incidence and emergence angles over the planetary disk. That said, it is not clear why the 1000~nm / 800~nm reflectance ratio for the sprinkled basalt powder behaves differently from the other two, but it might be due to a spurious drift in the lamp output (Fig~\ref{fig:stability}).

\subsubsection{Phase Reddening}
\label{sec:phase_reddening}

\begin{figure}
\centering
\includegraphics[width=8cm,angle=0]{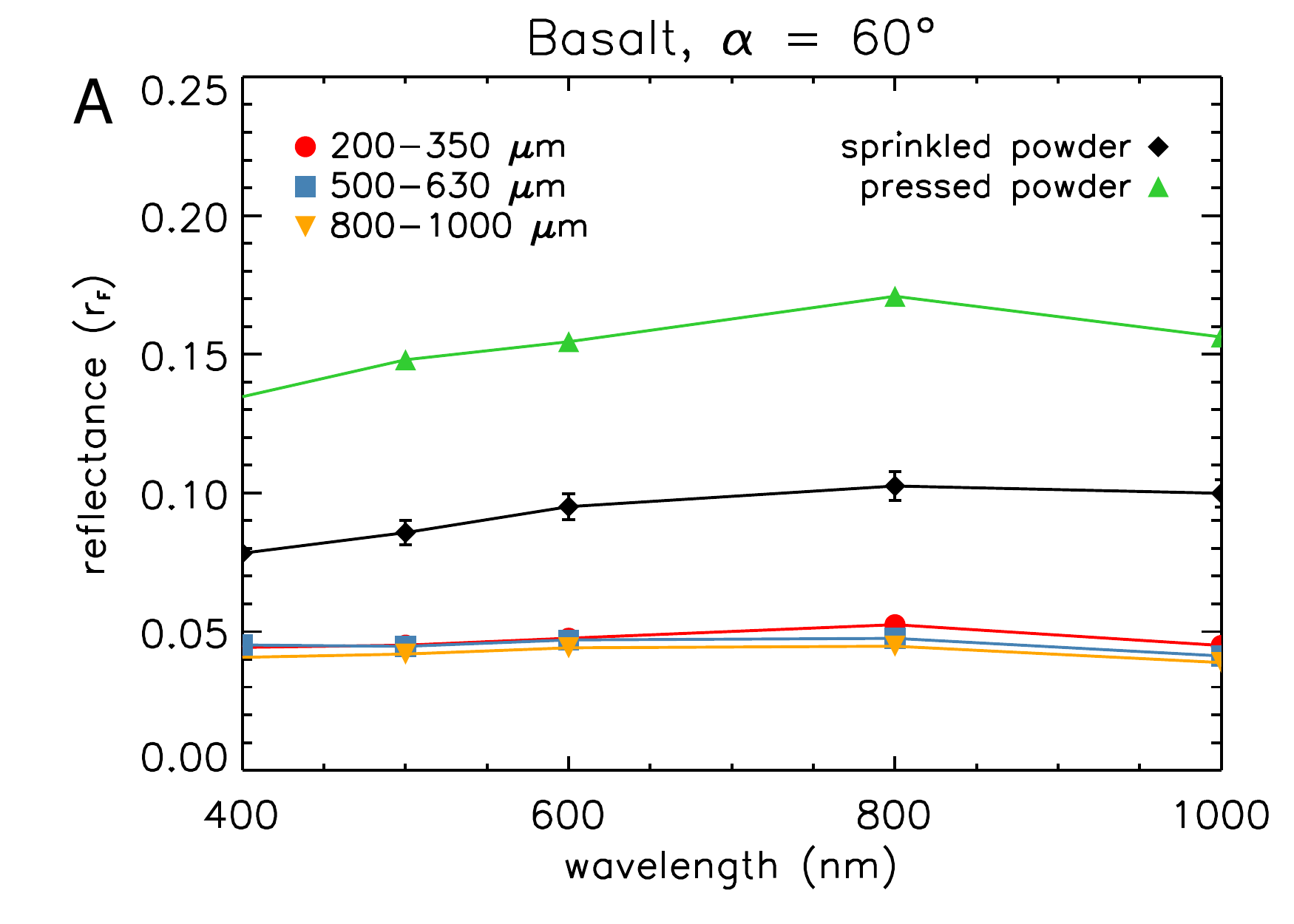}
\includegraphics[width=8cm,angle=0]{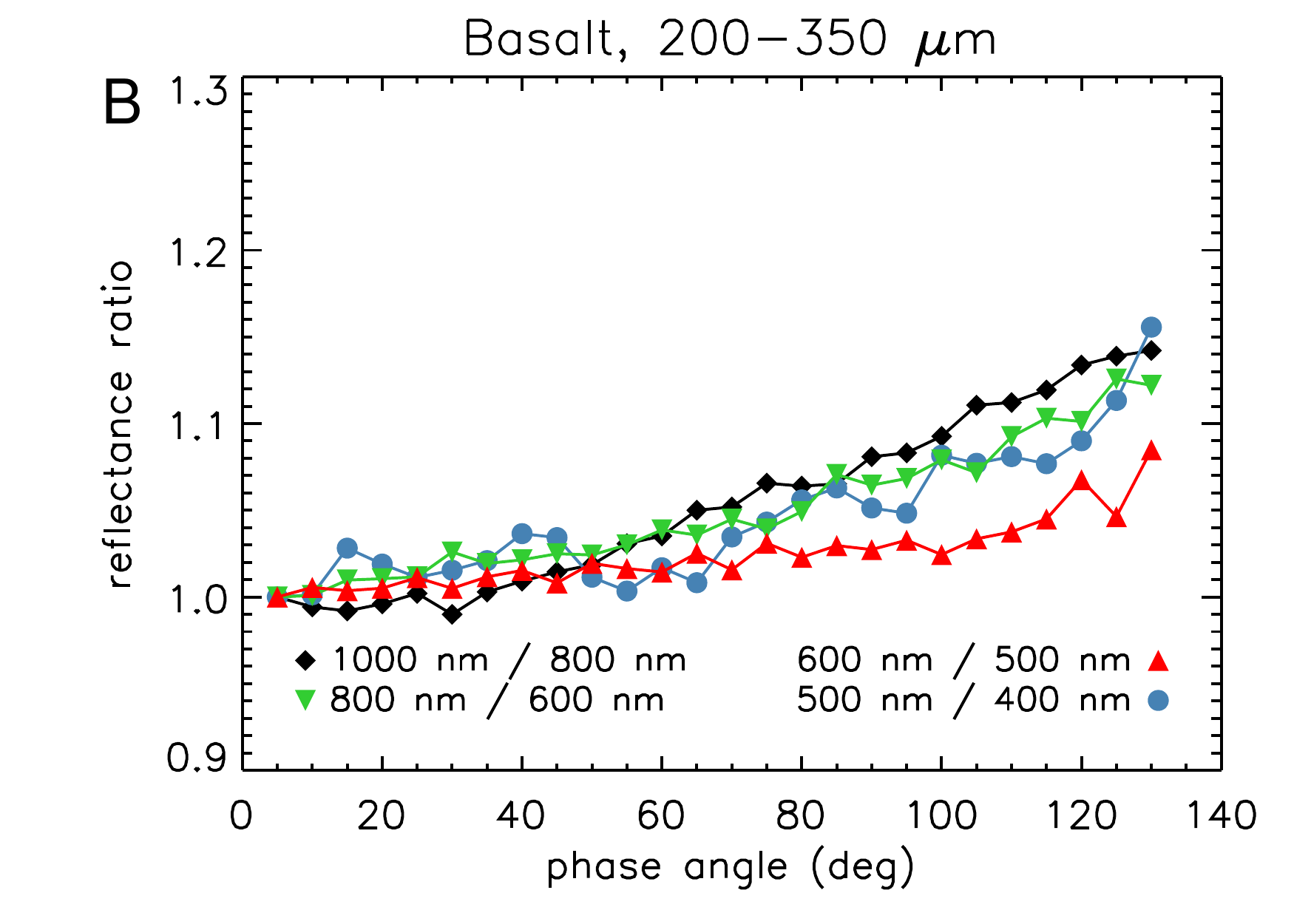}
\includegraphics[width=8cm,angle=0]{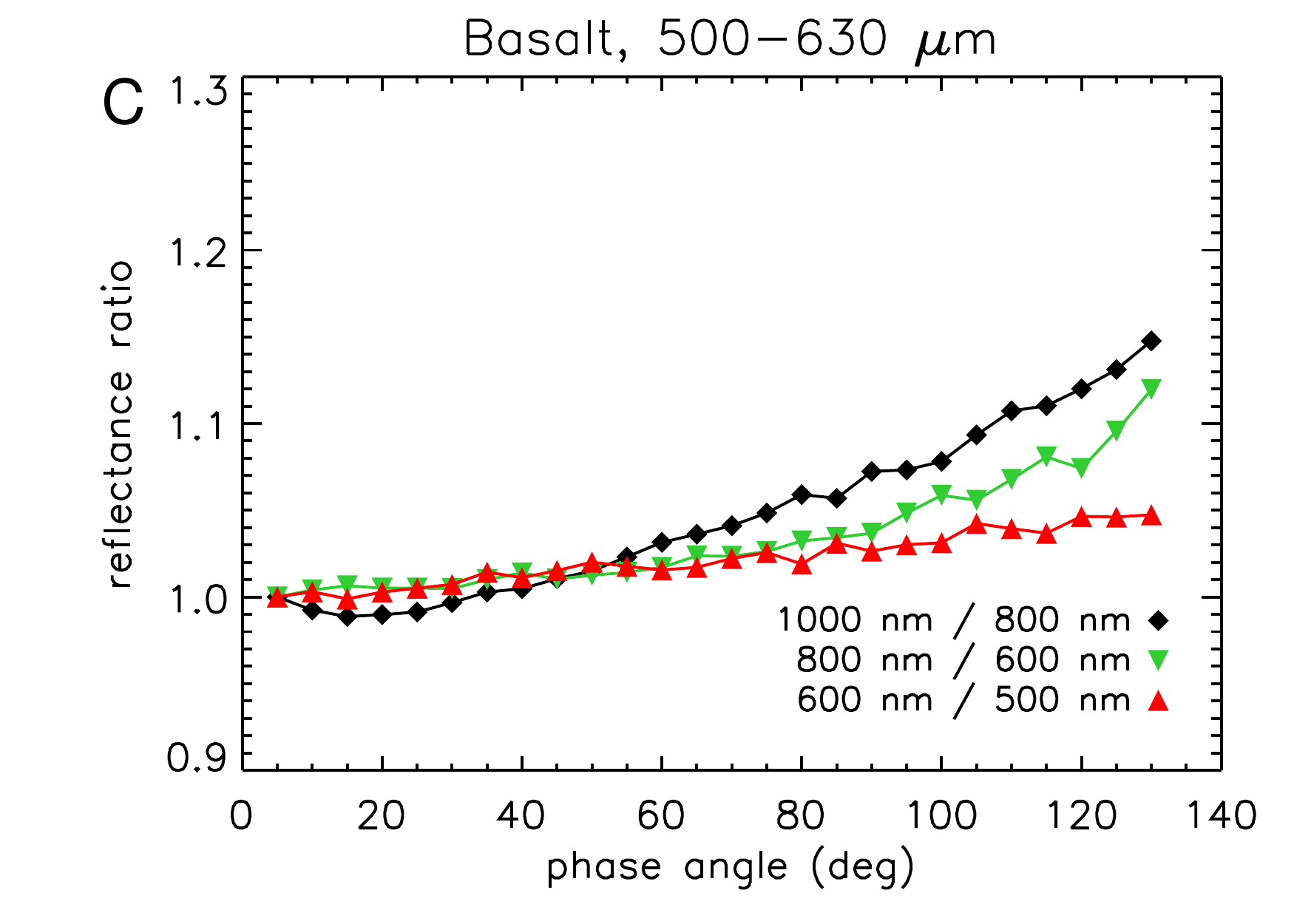}
\includegraphics[width=8cm,angle=0]{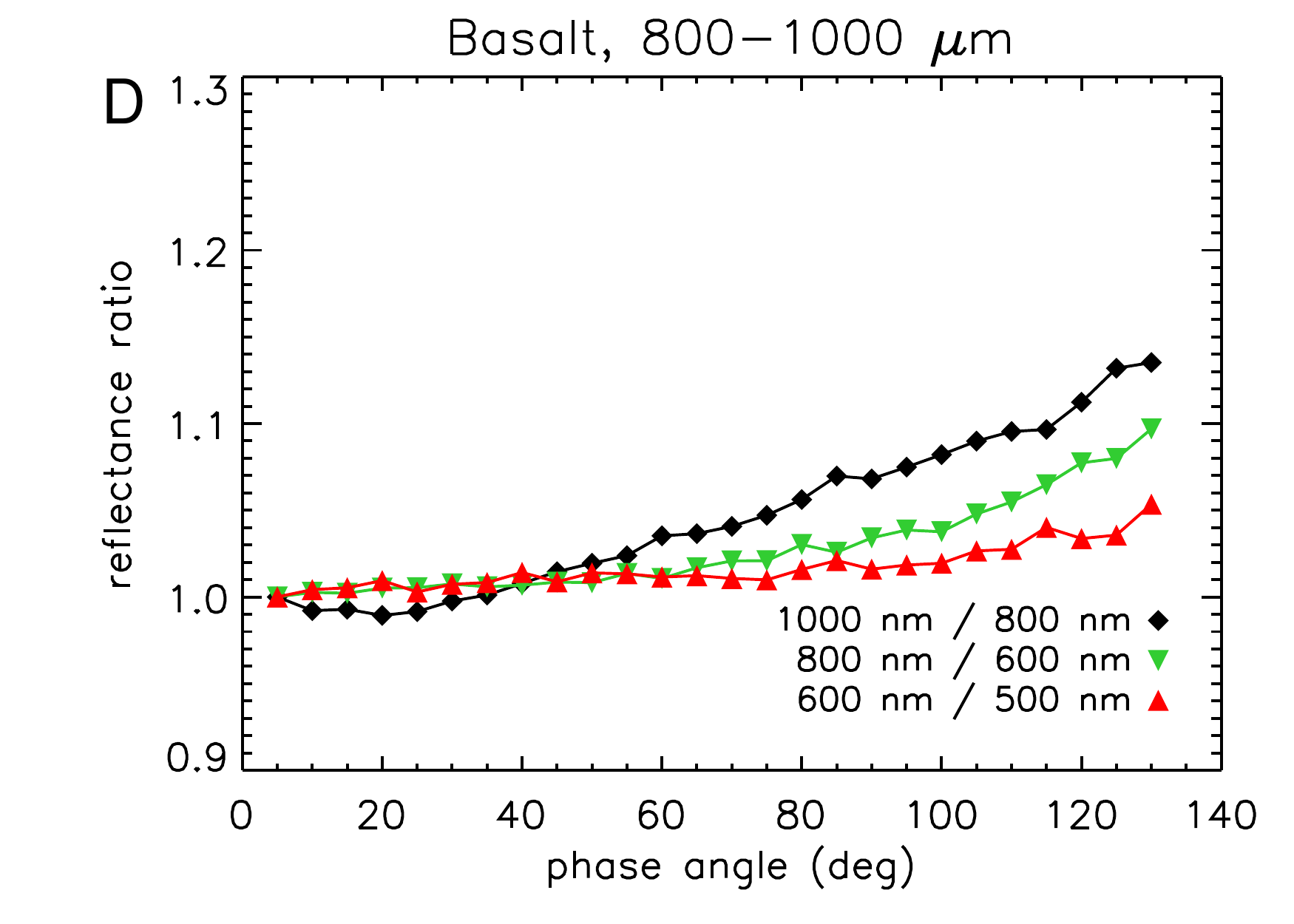}
\includegraphics[width=8cm,angle=0]{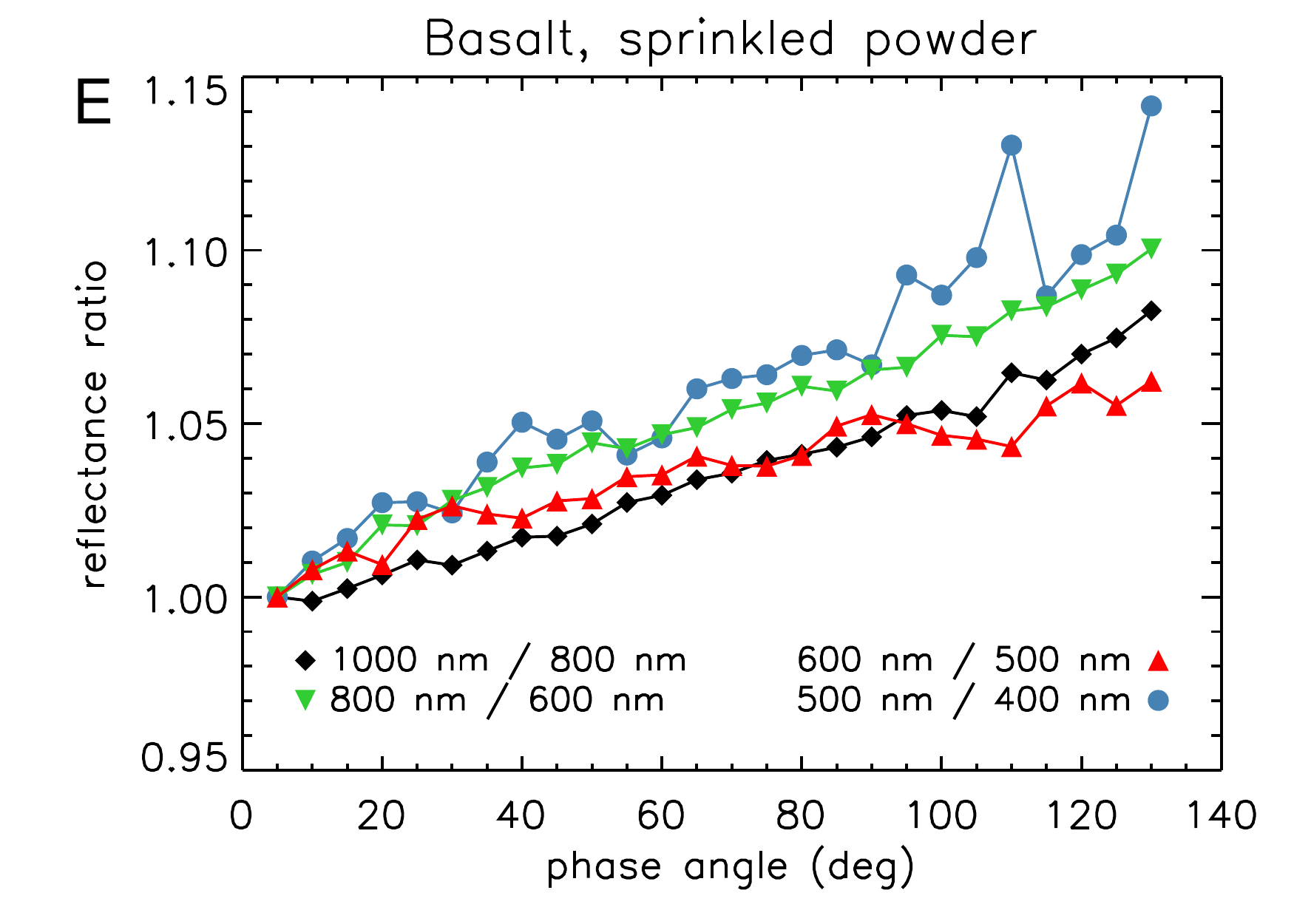}
\includegraphics[width=8cm,angle=0]{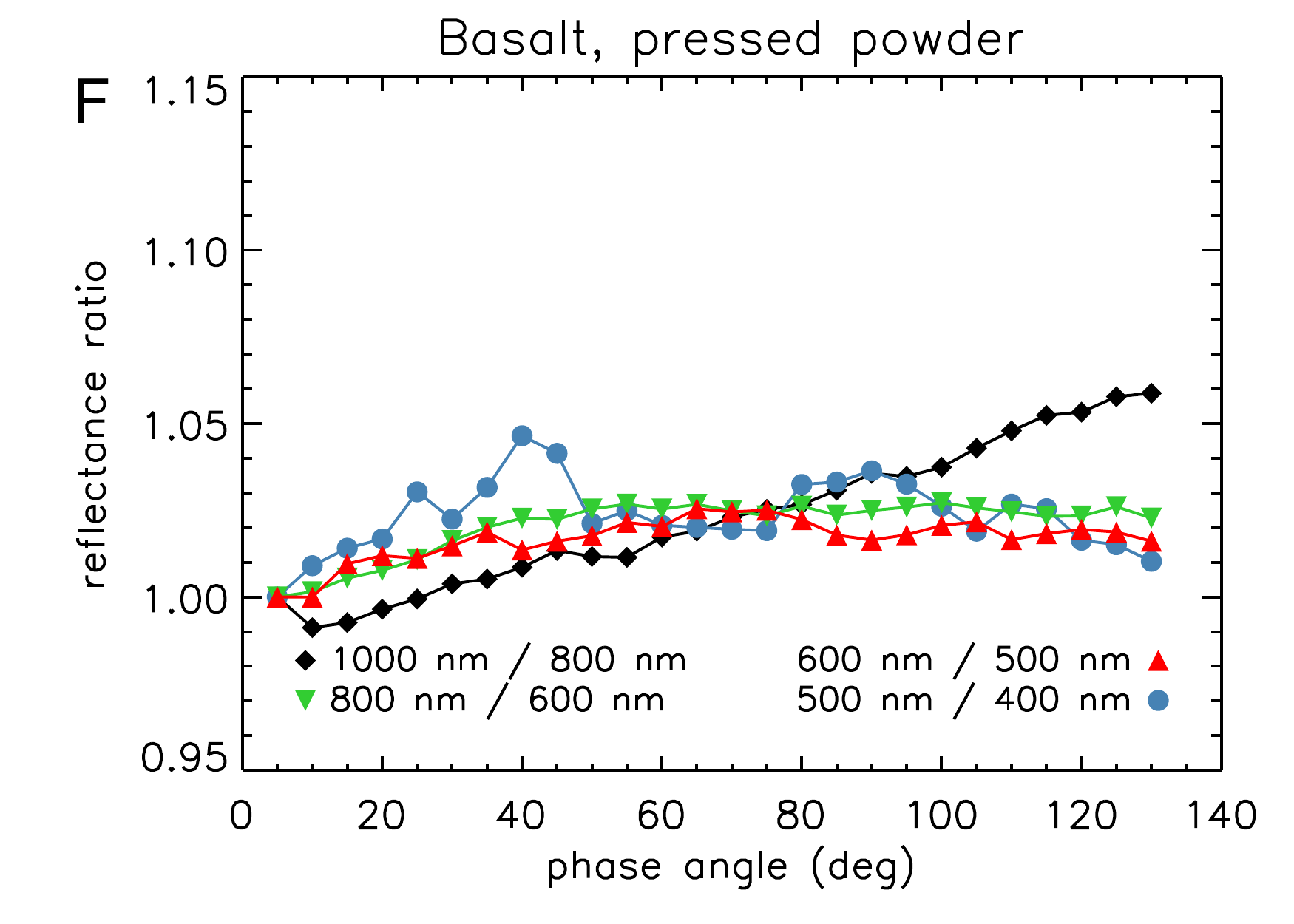}
\caption{Phase reddening for the basalt samples. {\bf A}:~Spectrum at $\iota = 60^\circ$ and $\epsilon = 0^\circ$ ($\alpha = 60^\circ$). Typical uncertainty is 5\% for each data point (shown for only one data set for clarity). {\bf B-F}: Phase reddening for the different particle sizes ($\iota = 60^\circ$, normalized at $\alpha = 5^\circ$). Note the different scale of the powder reflectance ratios.}
\label{fig:basalt_phase_reddening}
\end{figure}

\begin{figure}
\centering
\includegraphics[width=8cm,angle=0]{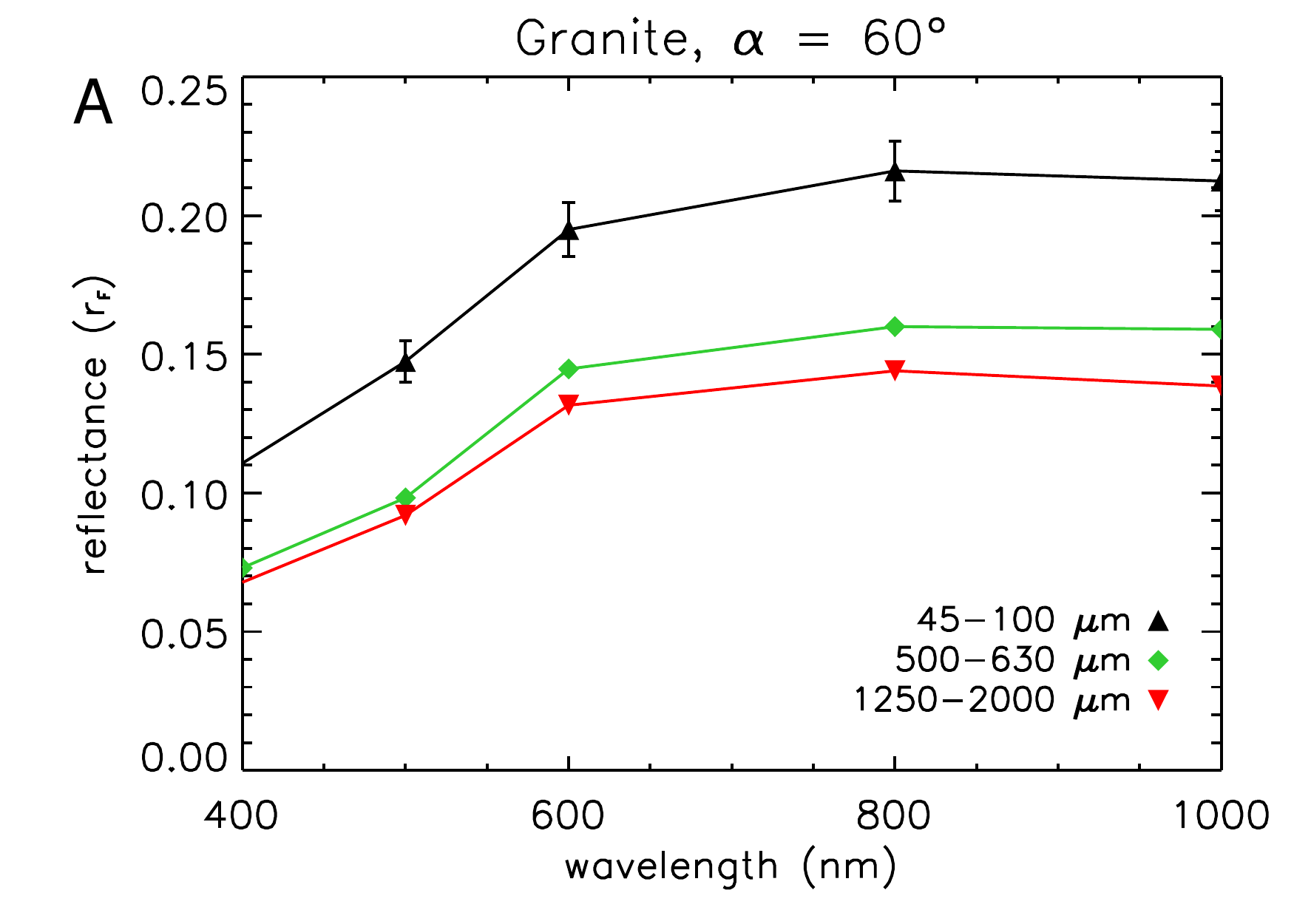}
\includegraphics[width=8cm,angle=0]{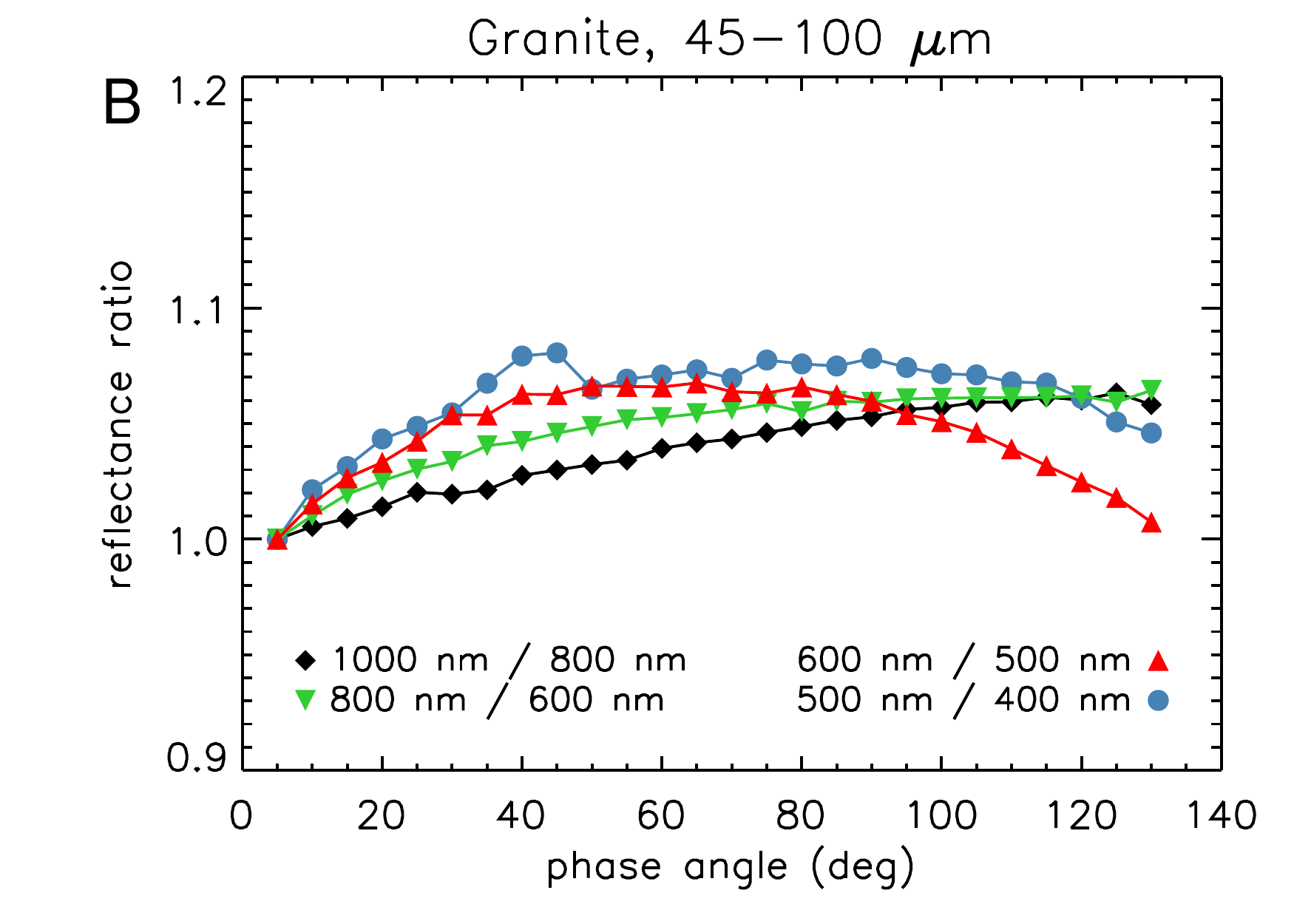}
\includegraphics[width=8cm,angle=0]{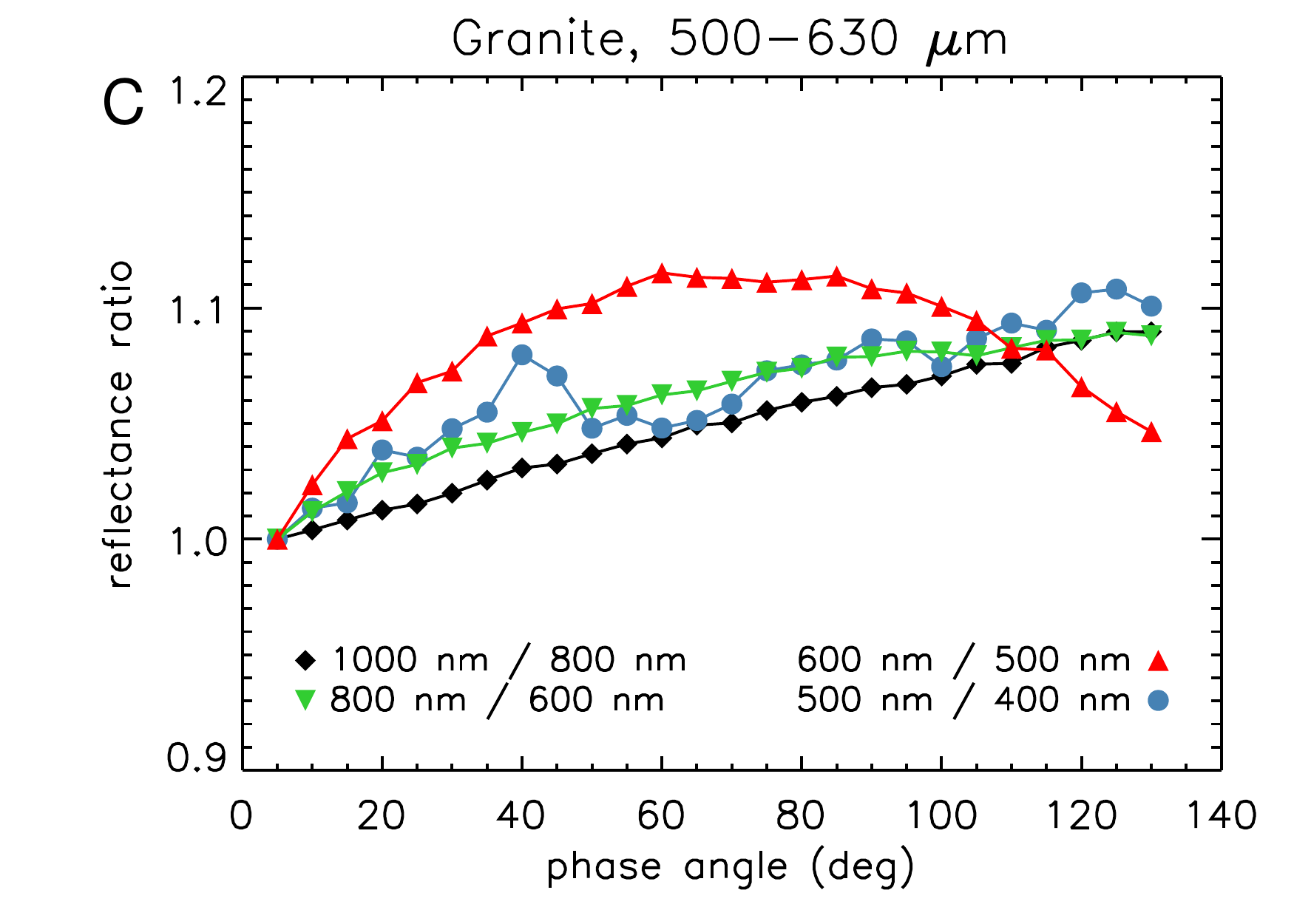}
\includegraphics[width=8cm,angle=0]{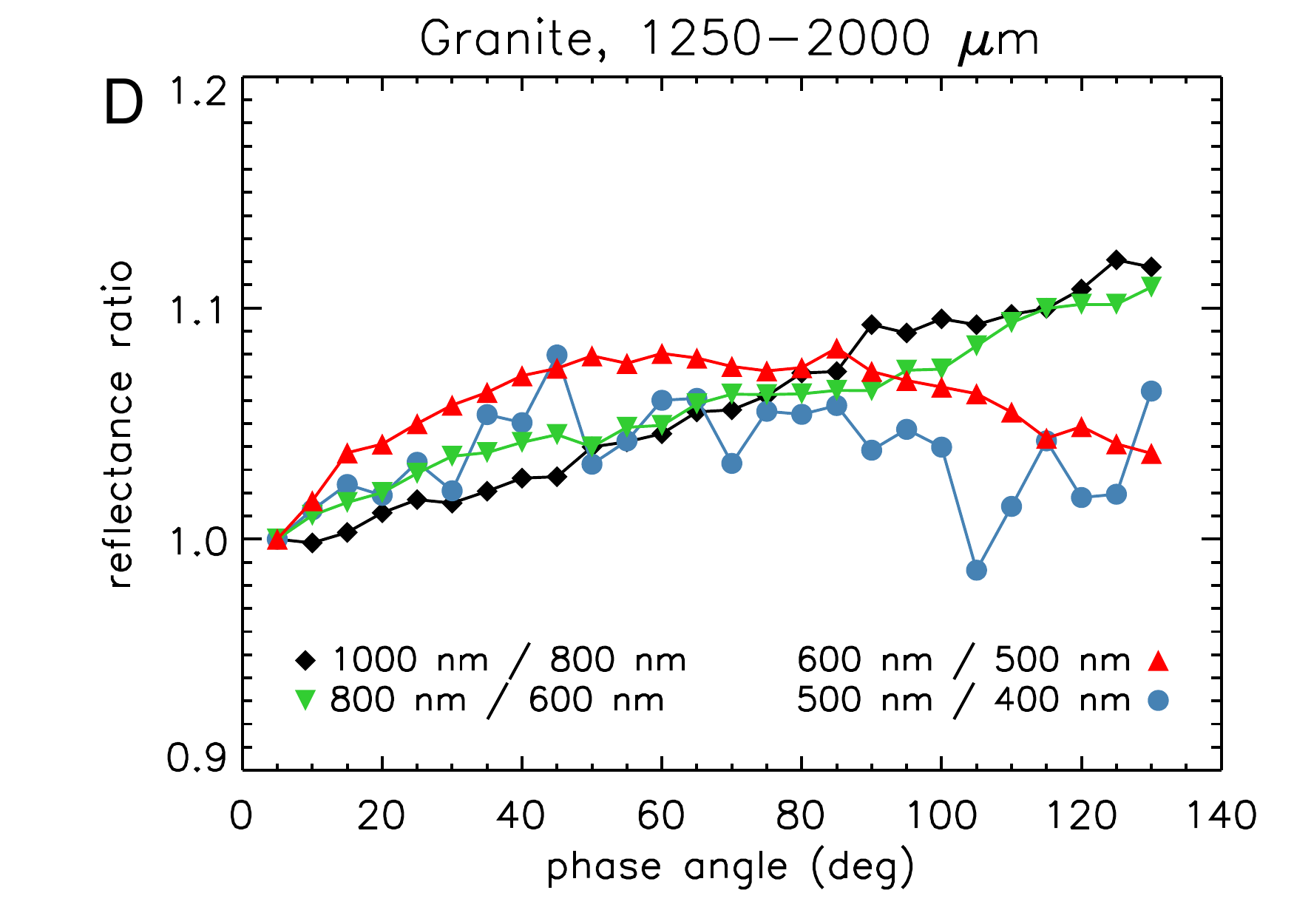}
\caption{Phase reddening for the granite samples. {\bf A}:~Spectrum at $\iota = 60^\circ$ and $\epsilon = 0^\circ$ ($\alpha = 60^\circ$). Typical uncertainty is 5\% for each data point (shown for only one data set for clarity). {\bf B-D}: Phase reddening for the different particle sizes ($\iota = 60^\circ$, normalized at $\alpha = 5^\circ$).}
\label{fig:granite_phase_reddening}
\end{figure}

\begin{figure}
\centering
\includegraphics[width=8cm,angle=0]{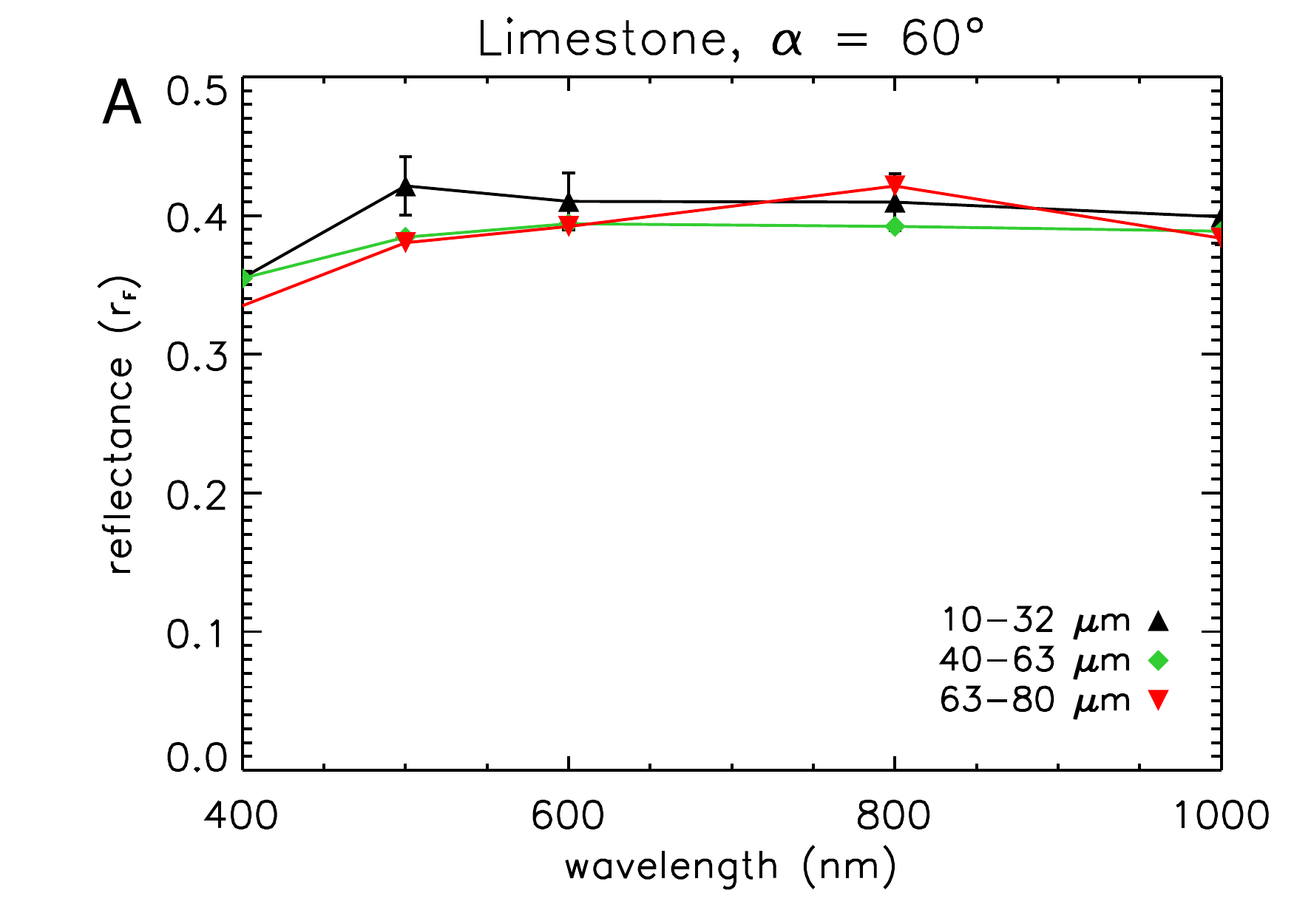}
\includegraphics[width=8cm,angle=0]{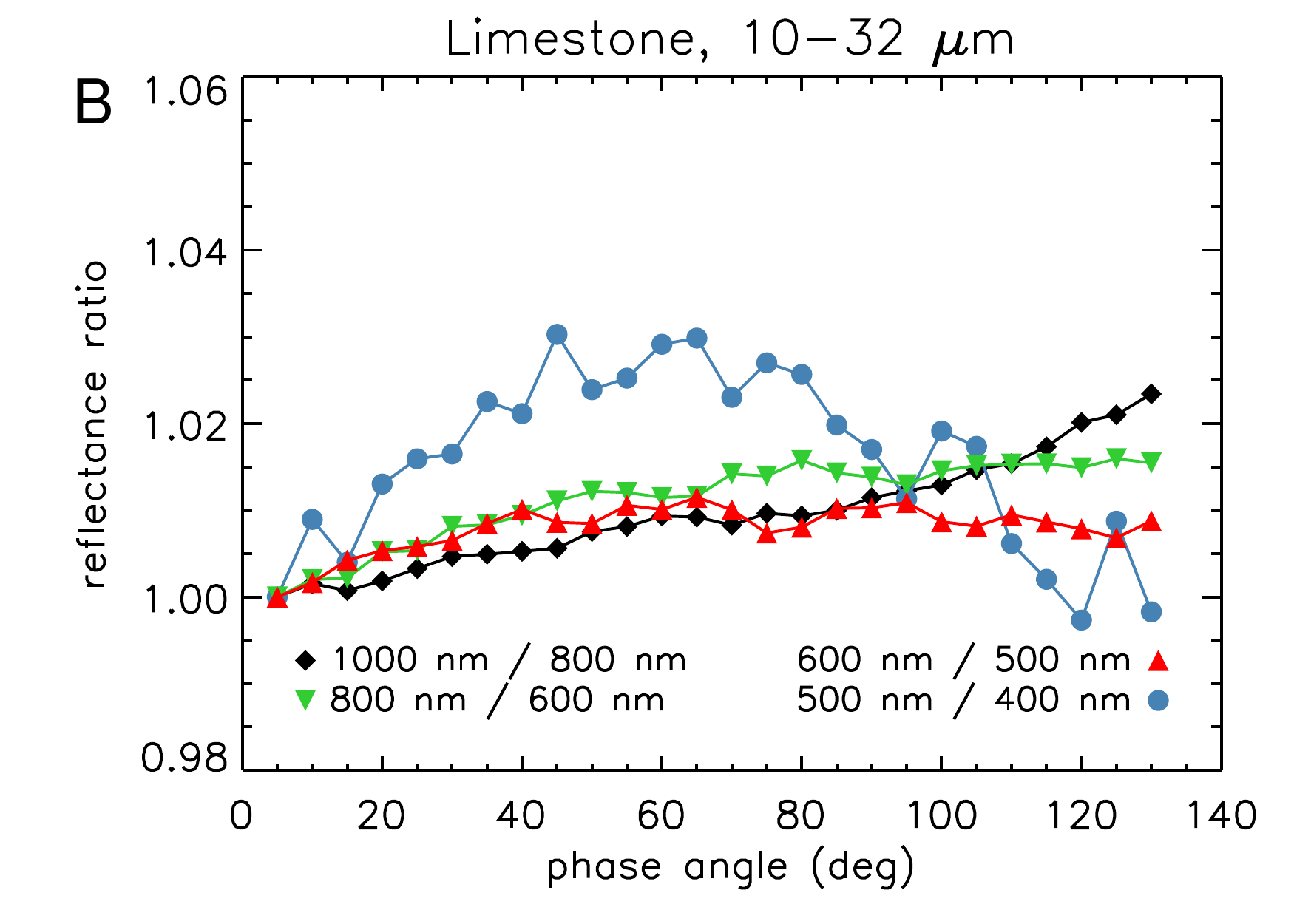}
\includegraphics[width=8cm,angle=0]{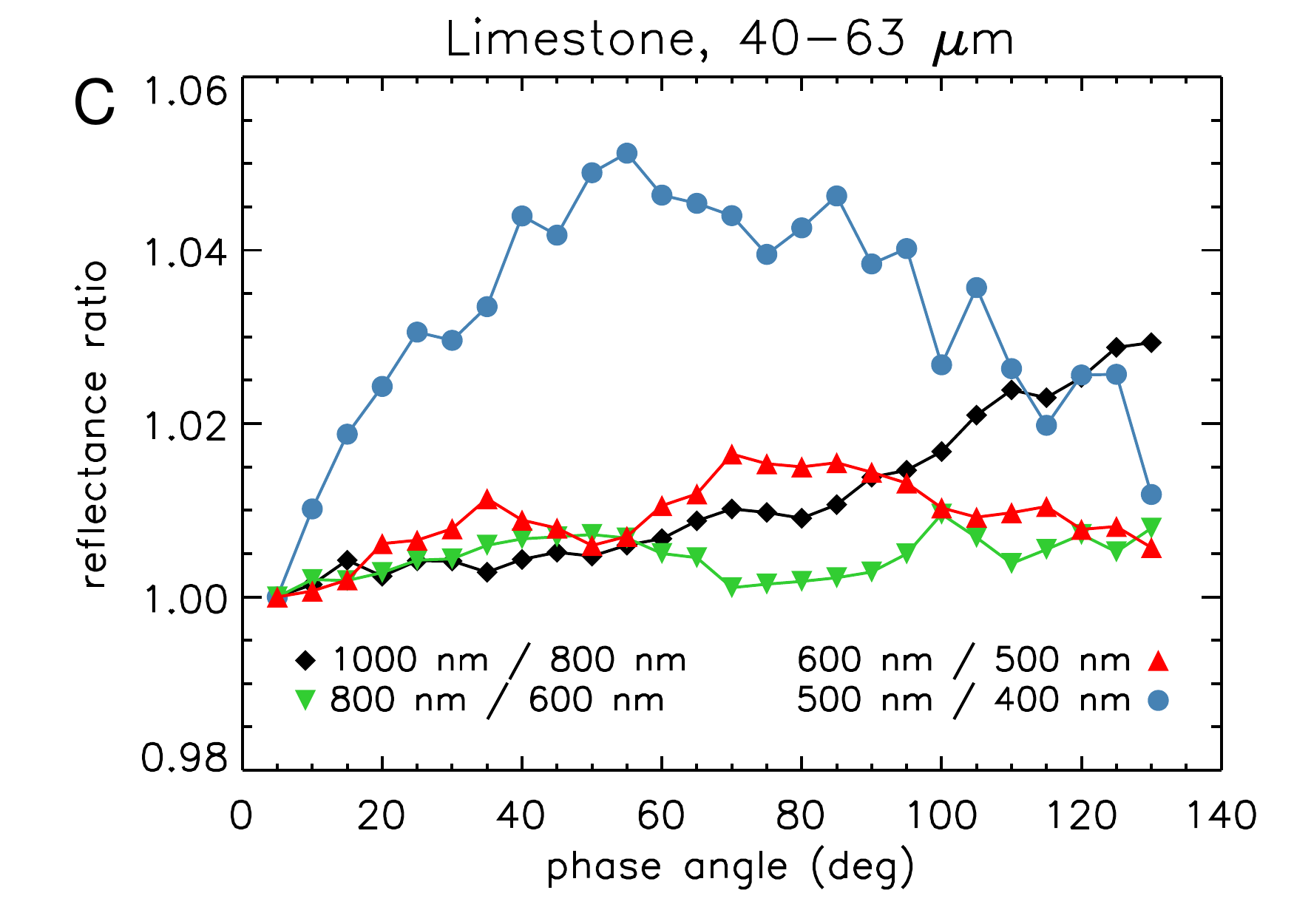}
\includegraphics[width=8cm,angle=0]{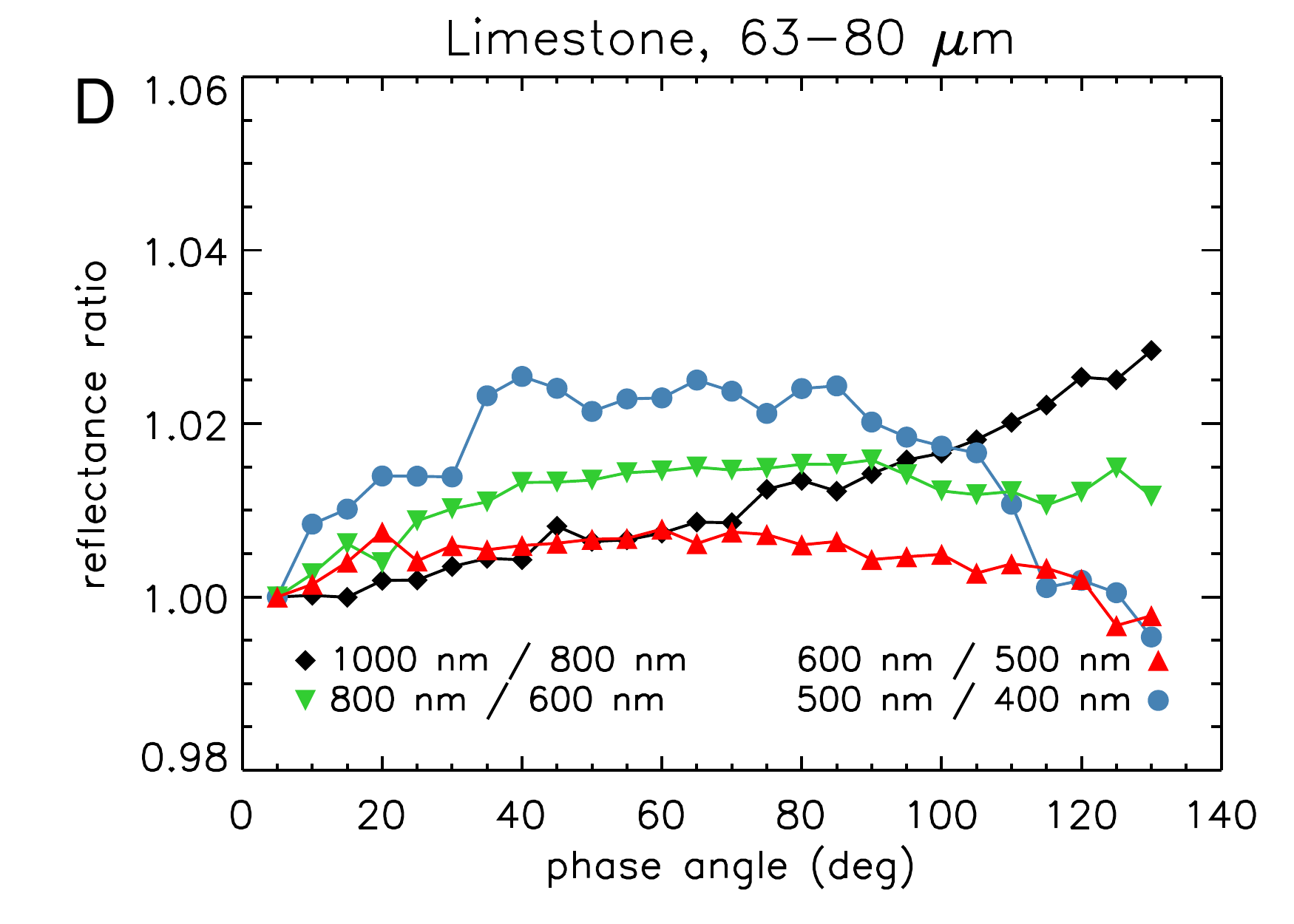}
\caption{Phase reddening for the limestone samples. {\bf A}:~Spectrum at $\iota = 60^\circ$ and $\epsilon = 0^\circ$ ($\alpha = 60^\circ$). Typical uncertainty is 5\% for each data point (shown for only one data set for clarity). {\bf B-D}: Phase reddening for the different particle sizes ($\iota = 60^\circ$, normalized at $\alpha = 5^\circ$). The {\bf B} and {\bf D} samples were shaken to flatten the surface, the {\bf C} sample was sprinkled.}
\label{fig:lime_stone_phase_reddening}
\end{figure}

We quantify phase reddening using the $\iota = 60^\circ$ phase curves. These were acquired by keeping the lamp arm at $\iota = 60^\circ$, and varying the detector arm angle from $\epsilon = 70^\circ$ to $-70^\circ$ in steps of $5^\circ$ (negative values are on the opposite side of the lamp arm). The $\epsilon = 55^\circ$ to $-70^\circ$ measurements then correspond to the phase angle range of $\alpha = 5^\circ$ to $130^\circ$. Reflectance ratios were constructed by dividing phase curves acquired at different wavelengths. We consider four reflectance ratios: 1000~nm / 800~nm, 800~nm / 600~nm, 600~nm / 500~nm, and 500~nm / 400~nm. The color changes observed for the basalt, granite, and limestone samples are shown in Figs.~\ref{fig:basalt_phase_reddening}, \ref{fig:granite_phase_reddening}, and \ref{fig:lime_stone_phase_reddening}, respectively. Also shown is a reflectance spectrum of each material at a geometry of $\iota = 60^\circ$ and $\epsilon = 0^\circ$ ($\alpha = 60^\circ$). The ratios are normalized at the smallest phase angle of observation ($5^\circ$) to facilitate a comparison of the reddening behavior for the different ratios.

The reflectance ratio profiles exhibit a variety of shapes: a monotonic increase, an increase at low phase angles and decrease at high angles, and the inverse of the latter. The profile shape appears to depend on at least three factors. The first is the local slope of the reflectance spectrum. If the reflectance increases from the small to the large wavelength in a particular ratio, the reflectance ratio as a function of phase angle increases at small phase angle (reddening), and decreases at large phase angle (bluing). We refer to this as the ``arch'' shape. The arch is especially well visible for the 600~nm / 500~ nm ratio of the granite samples (Fig.~\ref{fig:granite_phase_reddening}, red symbols), and the 500~nm / 400~ nm ratio of the limestone samples (Fig.~\ref{fig:lime_stone_phase_reddening}, blue symbols). The stronger the increase in reflectance with wavelength, the larger the amplitude of the arch. For the limestone samples, the arch associated with the 40-63~\textmu m sample has the largest amplitude. This is related to the method of deposition into the petri dish. It was difficult to achieve a flat surface, so we tried both shaking the dish and sprinkling the sample. The 40-63~\textmu m sample was sprinkled, whereas the other two samples were shaken. We observed that the surface after shaking was slightly lumpy, and less flat than the sprinkled surface. Possibly, this depressed the arch.

\begin{figure}
\centering
\includegraphics[width=8cm,angle=0]{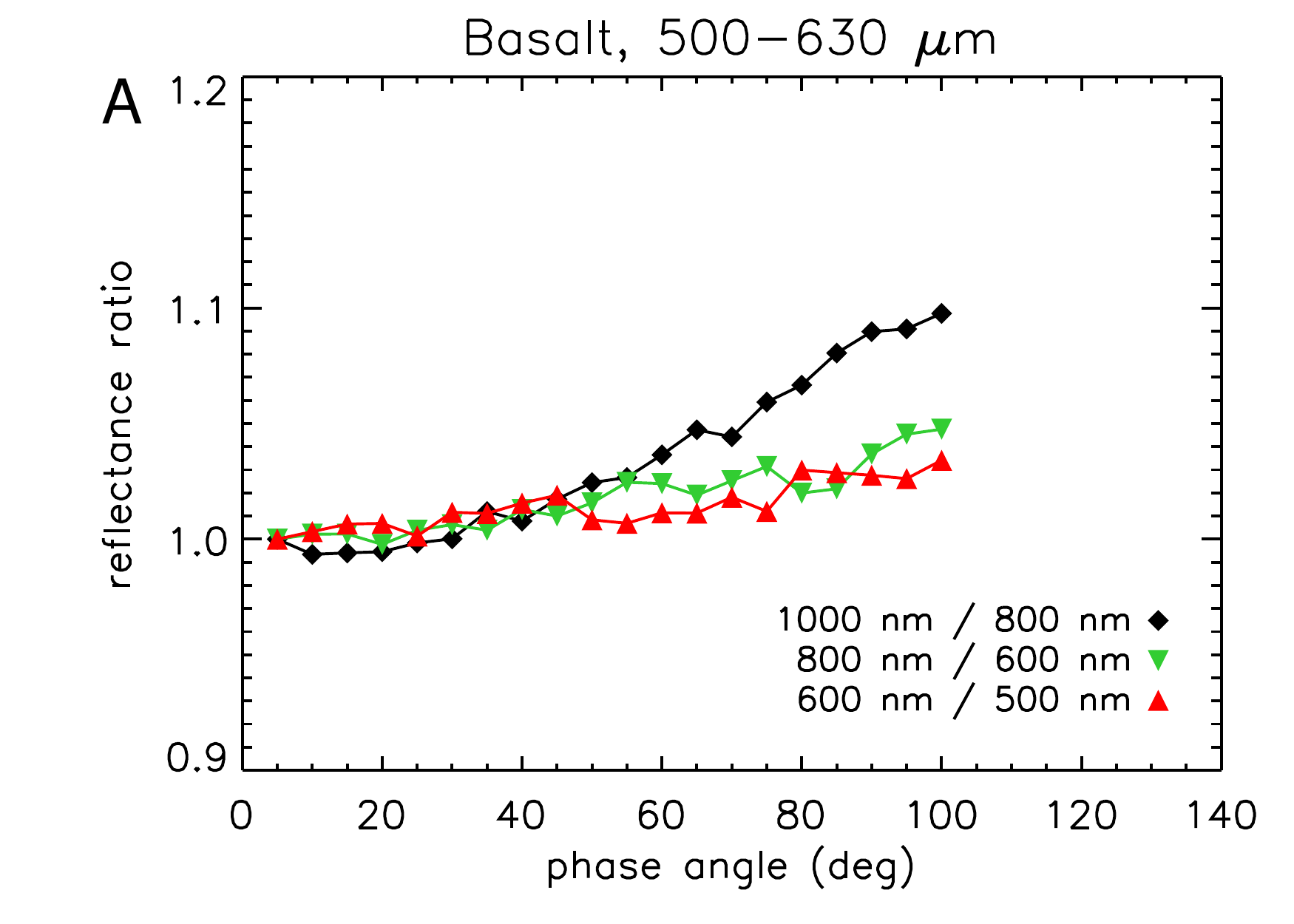}
\includegraphics[width=8cm,angle=0]{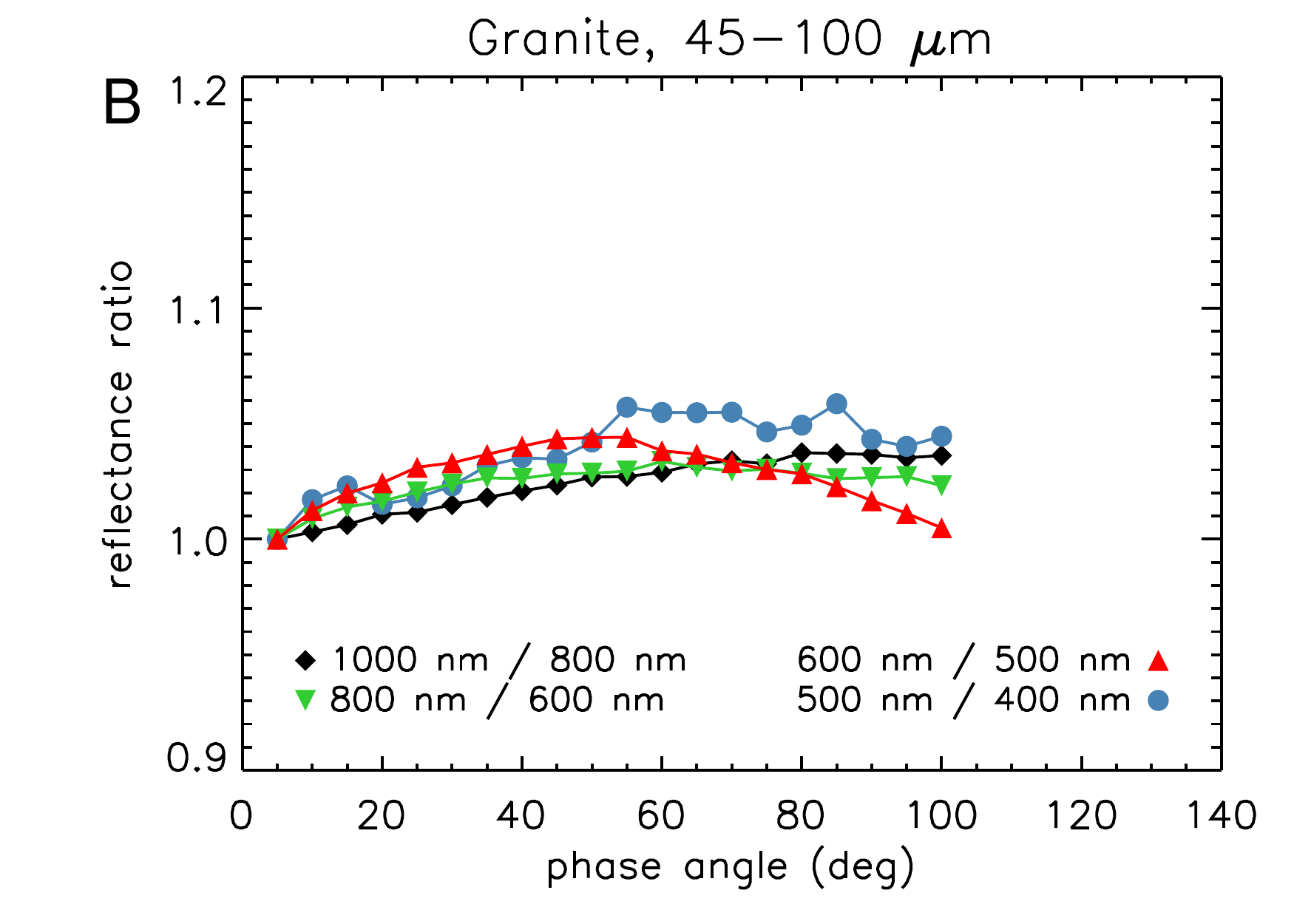}
\caption{Phase reddening for basalt ({\bf A}) and granite ({\bf B}) samples at $\iota = 30^\circ$ (normalized at $\alpha = 5^\circ$).}
\label{fig:phase_reddening_30deg}
\end{figure}

The phase angle at the maximum of the arch, where reddening turns to bluing, depends on the incidence angle. Reflectance ratio measurements of granite with $\iota = 30^\circ$ (Fig.~\ref{fig:phase_reddening_30deg}) have the phase angle at the arch maximum at roughly half the $\iota = 60^\circ$ value, revealing the role of geometry in determining the arch shape. We cannot verify the existence of an inverse arch, as our samples do not include a material with a clear decrease in reflectance in the 400-1000~nm wavelength range. Only for the 1000~nm / 800~nm ratio of the basalt samples (except the sprinkled powder) we observe a slight bluing at small phase angles, which is quickly overtaken by reddening. This bluing may indeed result from a local decrease in the reflectance spectrum (our measurements do not have the required accuracy). But since it is observed for the large basalt particles, which are fully opaque, it must be introduced by the surface of the particles. The micrographs in Fig.~\ref{fig:samples} show their surface to be extremely rough. The arch shape only appears if the particles are semi-transparent. For example, in Fig.~\ref{fig:basalt_phase_reddening} it is seen for the 600~nm / 500~ nm ratio of the pressed basalt powder (of which the particles are semi-transparent), but not for the coarse basalt particles (which are completely opaque), consistent with an observation by \citet{P13}. As light is blocked completely inside large basalt particles, the reflected light is not affected by the index of refraction, and the reflectance spectrum does not increase from 500 to 600~nm.

While a weak arch shape is present for the pressed basalt powder (except for the 1000~nm / 800~nm ratio), it is absent for the sprinkled basalt powder, where it is replaced by monotonic reddening (Fig.~\ref{fig:basalt_phase_reddening}E and F). This leads us to the second factor affecting the shape of the reflectance ratio profile: surface roughness on the scale of the particle size. The pressed and sprinkled basalt sample are exactly the same, except that the former is essentially smooth on the scale of the (largest) particle size, and the latter features the fairy castles. These towering structures (as seen under the microscope) change the photometric properties completely. They not only change the surface from predominantly forward scattering to back-scattering (Fig.~\ref{fig:basalt_phase_curves}), they also repress the reflectance ratio arch and introduce monotonic reddening at all ratios selected for our experiment.

The third factor affecting the shape of the reflectance ratio curve is simply wavelength. All samples exhibit monotonic reddening with increasing phase angle that is stronger for ratios of larger wavelengths. The 1000~nm / 800~nm ratio curve is fully dominated by strong, monotonic reddening, even though the spectrum is often flat in this wavelength range. Reddening for the 800~nm / 600~nm ratio is weaker, but still significant. Looking at the granite curve associated with this ratio in Fig.~\ref{fig:granite_phase_reddening}B, we expect to see a shallow arch, since the spectrum increases slightly from 600 to 800~nm. However, while the ratio increases for small phase angles, it does not downturn at larger angles. But this is exactly the shape one would expect if the arch is superposed on a linear increase. This increase represents monotonic reddening at a lower rate than that for the 1000~nm / 800~nm ratio.

\section{Simulation}
\label{sec:simulation}

\subsection{Method}

In order to understand the physics behind phase reddening, both the observed reddening and bluing, we performed ray-tracing simulations. The model that we use has been applied to the analysis of the reflective properties of powder-like surfaces, as reported in spectroscopic and polarization studies by \citet{GS02,GS03a,GS03b} and \citet{SG05}. The model is described in detail by \citet{GS08}. In short, it works as follows. A particulate medium of randomly shaped particles is generated, each represented by a polyhedron with triangular facets. It is illuminated by a large number of rays that are reflected, refracted, and partly absorbed on their way through the particles. The rays are traced from their first to their last interaction with the surfaces of the various particles. The interactions are described by means of the Fresnel equations and Snell's law. The surfaces of the particles have curvature radii large enough for the Fresnel formulas to apply. The model medium is characterized by the volume fraction (or packing density) of particles $\rho$ and the average particle size $d$. The simulations were done for two materials: basalt and granite. The model media are generated with nearly constant particle sizes of 10, 300, 600 and 1000~\textmu m, i.e.\ with a narrow size distribution to match that of the experimental samples. The 10~\textmu m sample represents the case of highly transparent particles, for which multiple scattering is important. Light that emerges from the surface after multiple scattering events has an angular distribution that is almost isotropic (diffuse). The materials are described by the complex index of refraction $m(\lambda) = n + i\kappa(\lambda)$, with refractive index $n$, extinction coefficient $\kappa$, and wavelength $\lambda$. The real part $n$ is taken to be constant, as it very weakly depends on the wavelength in the range of our experiment \citep{L98}. The wavelength dependence of the imaginary part $\kappa$ for basalt was adopted from \citet{P73}. We could not locate optical constants for granite, so we estimated these from our own reflectance measurements, assuming a constant composition using the approach of \citet{R07}.

The packing density (defined as the volume occupied by the particles relative to the total volume) of the simulated soil is $\rho = 0.3$. This density was reached after evolving the system through randomly wobbling and inflating particles. Real powders have packing densities around 0.5, but to achieve such values one needs to model gravity and elastic interaction of the particles. While $\rho = 0.3$ is an underestimation, the packing density appears to only weakly affect the shape of reflectance spectra \citep{P91,S99}. Note that the packing density of the simulated surface should not be seen as equal to the porosity of the fairy castle surface. The simulated surface has voids on the scale of the particles, and is smooth on the scale of the particles. The fairy castles are huge compared to the scale of the particles (Fig.~\ref{fig:basalt_powder}). The packing density may have a certain typical value inside the castles, but is zero outside. Therefore, the simulations are only representative of surfaces that are smooth on the scale of the particles, like that of pressed basalt.

\subsection{Results}

\begin{figure}
\centering
\includegraphics[width=\textwidth,angle=0]{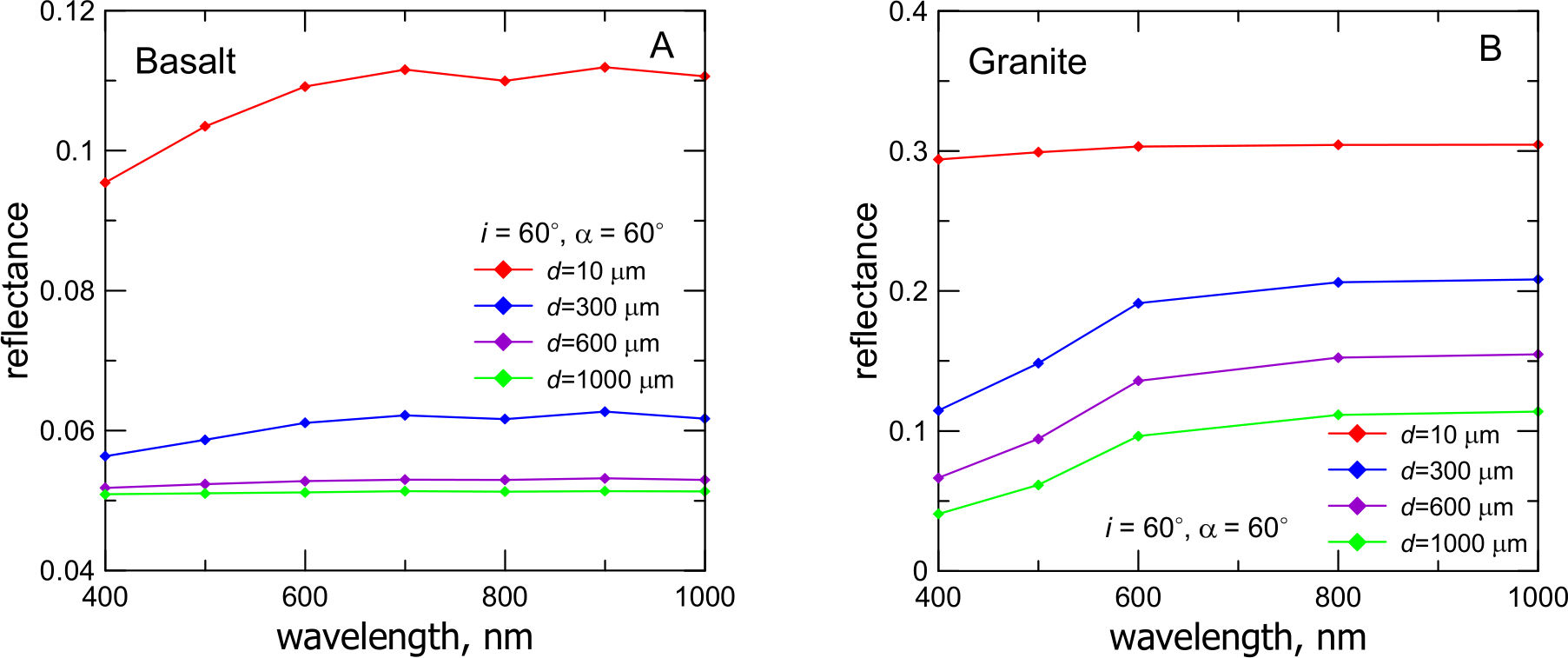}
\caption{Simulated reflectance spectra for different particle sizes at $\iota = 60^\circ$ and $\epsilon = 0^\circ$ ($\alpha = 60^\circ$). {\bf A}:~Basalt. {\bf B}:~Granite.}
\label{fig:sim_spectra}
\end{figure}

The simulations reproduce several aspects of our experiments. First, we consider the reflectance spectra of the different size fractions of basalt and granite as derived from the optical constants. The modeled reflectance is defined as the ratio of the number of rays collected at a certain phase angle and wavelength (and corresponding extinction coefficient $\kappa$) to the number of rays collected for totally transparent and non-absorbing particles ($\kappa = 0$) at the same geometry, multiplied by the cosine of the incidence angle. This makes the modeled reflectance equivalent to the radiance factor, and thereby directly comparable to the measurements. The model surfaces exhibit the observed property of brightening when the particles get smaller, as small particles are more transparent than large ones. The simulated spectra for basalt particles larger than 500~\textmu m are flat (compare Figs.~\ref{fig:basalt_phase_reddening}A and \ref{fig:sim_spectra}A), at the same absolute reflectance value as measured. The brightening of the model surface for 300~\textmu m particles is not observed, perhaps because the \citet{P73} optical constants are not fully representative for our samples (note that the simulated spectra have an absorption band at 800~nm that we did not observe). The model basalt powder should be compared with the pressed powder basalt sample, as the simulation assumes a flat, smooth surface. It is somewhat darker than observed, probably because the 10~\textmu m model particles are a little larger than those in the powder sample. The modeled granite spectra compare very well with the observations, which is reassuring given that the optical constants were derived from the measurements. For example, the modeled reflectance for 600~\textmu m particles (Fig.~\ref{fig:sim_spectra}B) is virtually identical to that measured for 500-630~\textmu m particles (Fig.~\ref{fig:granite_phase_reddening}A).

\begin{figure}
\centering
\includegraphics[width=\textwidth,angle=0]{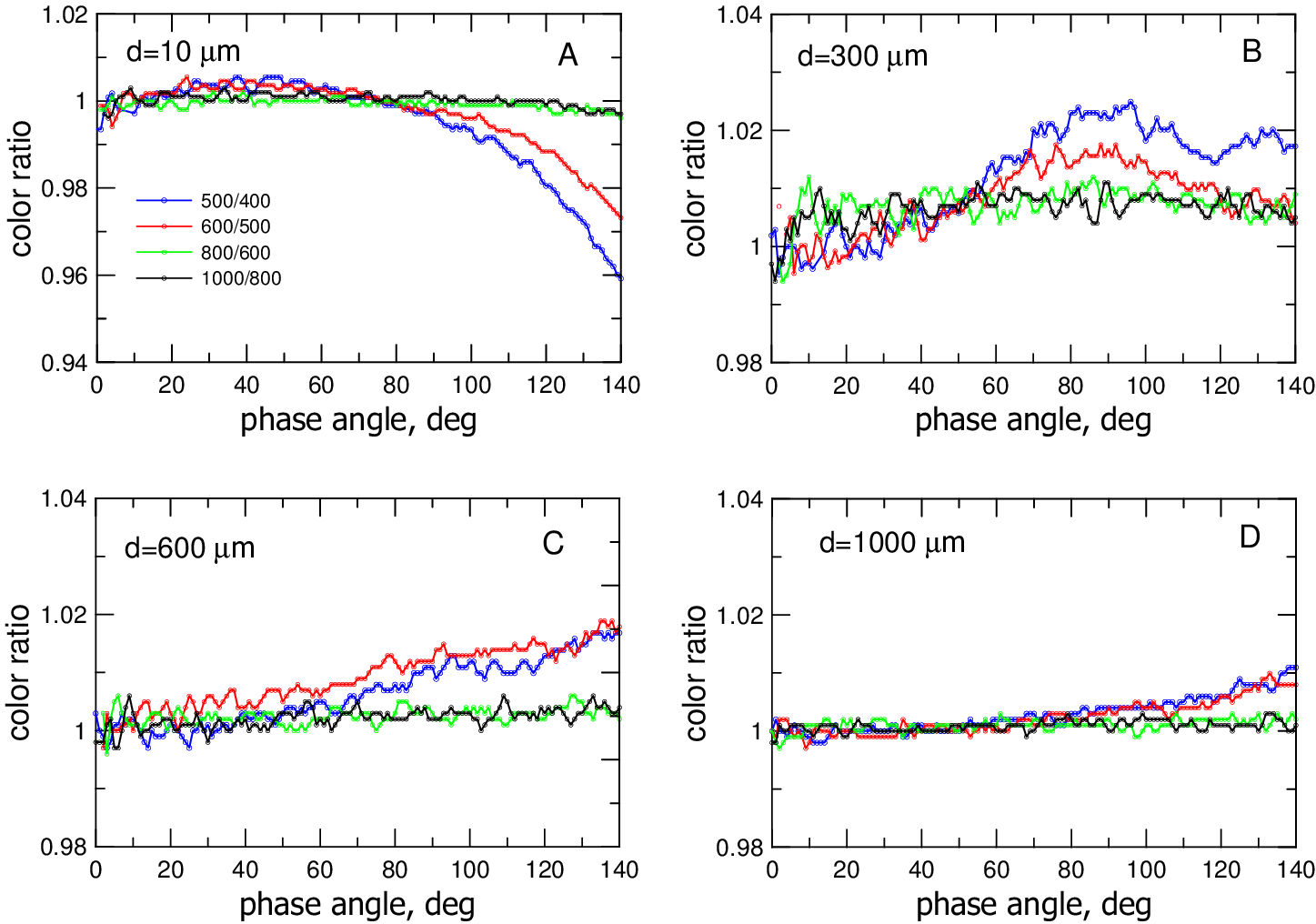}
\caption{Simulated reflectance ratios for basalt ($\iota = 60^\circ$ and $\epsilon = 0^\circ$, normalized at $\alpha = 5^\circ$), with particles sizes {\bf A}:~10~\textmu m, {\bf B}:~300~\textmu m, {\bf C}:~600~\textmu m, and {\bf D}:~1000~\textmu m.}
\label{fig:sim_col_ratios_basalt}
\end{figure}

\begin{figure}
\centering
\includegraphics[width=\textwidth,angle=0]{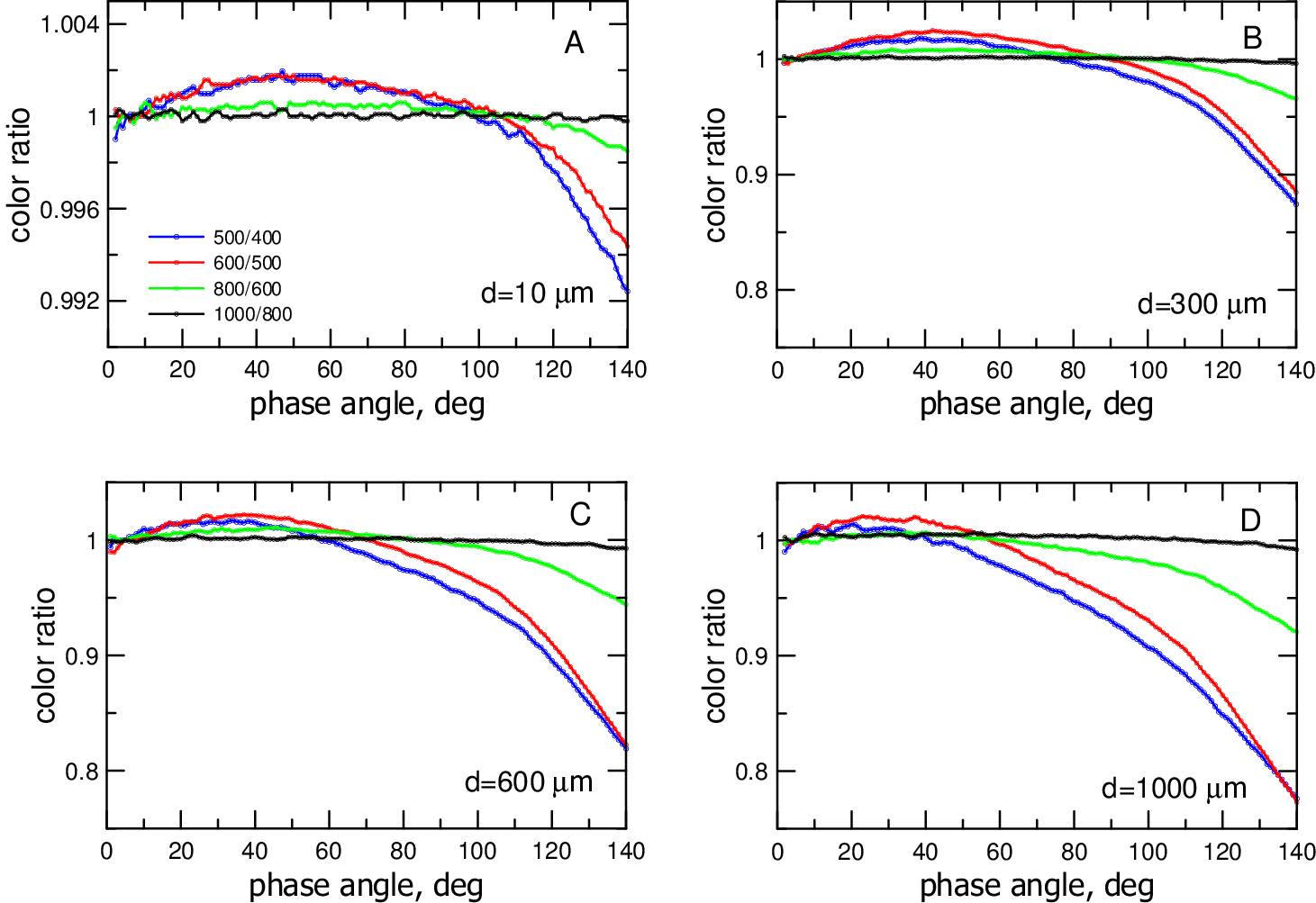}
\caption{Simulated reflectance ratios for granite ($\iota = 60^\circ$ and $\epsilon = 0^\circ$, normalized at $\alpha = 5^\circ$), with particles sizes {\bf A}:~10~\textmu m, {\bf B}:~300~\textmu m, {\bf C}:~600~\textmu m, and {\bf D}:~1000~\textmu m.}
\label{fig:sim_col_ratios_granite}
\end{figure}

The phase reddening exhibited by the model surfaces is displayed in Figs.~\ref{fig:sim_col_ratios_basalt} (basalt) and \ref{fig:sim_col_ratios_granite} (granite), as reflectance ratios equivalent to those measured in the experiments for four particle sizes. Only the smallest basalt particles (10~\textmu m) show the arch (reddening with increasing phase angle followed by bluing), and only for those wavelength ratios that are associated with an increase in the reflectance spectrum (500~\textmu m / 400~\textmu m and 600~\textmu m / 500~\textmu m). These particles are small enough to be semi-transparent, and multiple scattering in the medium leads to the arch shape of the reflectance ratios. The mechanism behind this was proposed by \citet{AF67} and further discussed by \citet{GS08}. The key parameter is the total path length that light rays have traveled inside the particles between the times of entrance and emergence from the surface. The simulations show that the path length averaged over many rays is non-monotonic with phase angle, showing a maximum for intermediate angles. Consider the red over blue reflectance ratio mentioned in the introduction. If the spectrum is flat, i.e.\ the absorption coefficient is the same for both red and blue light, the changing path length with phase angle is of no consequence for the color. If the spectrum is increasing, i.e.\ the absorption coefficient is lower for red light, the changing path length does affect the color; blue rays will be absorbed preferentially for the longer paths at intermediate phase angles, leading to the arch shape.

Basalt particles sized $>300$~\textmu m are essentially opaque. Most emergent rays have reflected off the particle surface only once, and do not ``feel'' the spectral absorption of the material. Yet for those reflectance ratios associated with a positive spectral slope, a weak monotonic reddening is observed. This is caused by rays that are refracted by the sharp edges of the particles. Only in the particle corners the ray path lengths are so short that rays can escape. Then, in case of a positive spectral slope, larger wavelengths are absorbed less. This phenomenon could be at least a partial explanation of the persistent reddening observed in our experiments. Fines clinging to the surface of the larger particles may act in a manner similar to sharp edges (the samples were dry-sieved). However, their role is difficult to assess; it is a very difficult task to simulate a large particle with fines on its surface, and to our knowledge, nobody has done this yet. The situation for the granite simulations is somewhat different, as granite has a much smaller imaginary part of the complex refractive index than basalt. The simulated surface of large particles now also exhibits the reddening arch for reflectance ratios associated with a positive spectral slope, in qualitative agreement with the experiments. However, the strong monotonic reddening for the higher reflectance ratios (800~\textmu m / 600~\textmu m and 1000~\textmu m / 800~\textmu m) we see  in the experiments is not reproduced by our simulations, neither for basalt nor granite. The amplitude of the simulated arch for granite is smaller than observed, which points at imperfections of the model. The geometric optics model with its faceted particles is not accurate if there are spatial scales of roughness on the order of the wavelength. Then the single scattering component of reflection from the particulate surface is not described properly. We suspect that is the case here, but a definite answer would require additional simulations using an accurate wave-optics method.

\begin{figure}
\centering
\includegraphics[width=\textwidth,angle=0]{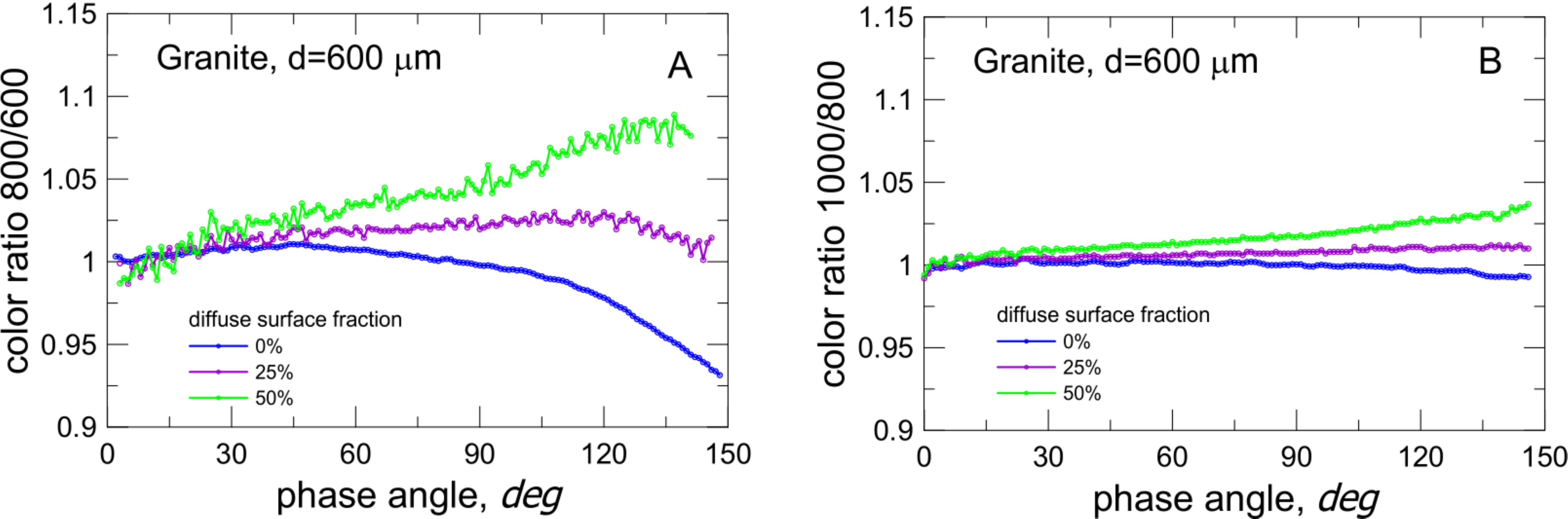}
\caption{Simulated reflectance ratios for a particulate granite surface with $d = 600$~\textmu m and different fractions of (diffuse) Lambertian elements on the particle surface to simulate roughness ($\iota = 60^\circ$ and $\epsilon = 0^\circ$, normalized at $\alpha = 5^\circ$). {\bf A}:~800~nm / 600~nm, {\bf B}:~1000~nm / 800~nm.}
\label{fig:sim_roughness}
\end{figure}

The mechanism responsible for monotonic reddening is probably a fundamental one, as we observe it for different materials and particle sizes. Probably, the particle morphology plays a significant role, particularly the roughness of the particle surface. \citet{S85,S87} showed analytically that in the Rayleigh-Rice approximation, roughness on the particle surface with a scale smaller than the wavelength can lead to spectral reddening. Another consequence of particle surface roughness is that it may lead to a marked drop of the specular reflectivity of the simulated sample \citep{GP09}. To test this reddening idea we performed simulations with a modified model, technically similar to that described by \citet{P07}. We use the concept of isotropically scattering (Lambertian) elements, randomly distributed over the surface of the particles of the medium \citep{M07}. Their reflectance is taken as proportional to $\lambda^{-4}$ (Rayleigh scattering). In this way we introduce an explicit wavelength dependence of the external reflection. Depending on the fraction of such elements with respect to the Fresnel facets, a certain proportion of the incident rays is reflected isotropically, simulating scattering resulting from roughness. So, in the modified model, we have now two components of local reflection: diffuse and specular, associated with the Lambertian elements and the Fresnel facets, respectively. The model spectra in Fig.~\ref{fig:sim_roughness} for granite become reddish with phase angle if the number of Lambertian elements is significant. Due to the $\lambda^{-4}$ dependence, the simulated reddening is weaker for the 1000~nm / 800~nm reflectance ratio than for the 800~nm / 600~nm ratio. However, our experiments reveal that the monotonic reddening is much stronger for ratios of larger wavelengths (1000~nm / 800~nm) than those of shorter wavelengths (600~nm / 500~nm). This suggests that at these wavelengths the relative scale of roughness is near the Rayleigh limit, where the surface is neither ideally smooth nor very rough. The Rayleigh condition implies that $h > \lambda \cos \iota / 8$, where $h$ is the root-mean-square height of the randomly rough surface (\citealt{T00}, p.~390). We conclude that the particles in our experiments are rough on a certain scale that results in strong reddening at larger wavelengths. In terms of the particle surface-roughness model, which, due to its simplicity, can only qualitatively demonstrate the effect of persistent reddening, this means that the fraction of diffuse surface elements increases with wavelength.

\section{Discussion}
\label{sec:discussion}

Our work reveals two fundamentally different types of spectral changes with phase angle observed for a surface of semi-transparent particles that is smooth on the scale of the particles. The first is the arch shape of the reflectance ratio (reddening at low and bluing at high phase angle), which appears if the reflectance spectrum has a positive slope from the lowest to the highest wavelength in the ratio. Our goniometer experiments and ray-tracing simulations reproduce the arch shape, which has been experimentally observed on many occasions \citep{AF67,GV82,K10,J13,P13}. The steeper the spectral slope, the larger the amplitude of the arch. We expect an inverted arch in case of a negative spectral slope, as was indeed observed by \citet{B12}. The simulations confirm the explanation in terms of average path length provided by \citet{AF67}, which is also consistent with the explanation offered by \citet{K10}. The second spectral change is a monotonic reddening that increases in strength with wavelength. As we observe these changes for the largest basalt particles, which are completely opaque, they must come from the surface of the particles. This was also suggested by \citet{B12}. Simulating particle surface roughness by including Lambertian surface elements in our model qualitatively reproduces the monotonic reddening. We hypothesize that the change in absorption band depth with phase angle, as observed for rocky solar system bodies \citep{C02,DV07,S12,L14} and in the laboratory \citep{P08}, is a combination of these two effects. This topic could be further explored with dedicated experiments.

The spectral changes are different when the surface is not smooth on the scale of the particles. Our experiments highlight the role of microscopic roughness (i.e.\ indistinct without the use of a microscope) in the form of structures that spontaneously form when sprinkling a powder: the \citet{HH63} fairy castles. Their presence on the surface of the basalt powder led to three major changes in the photometric properties: (1)~a steepening of the phase curve and suppression of the forward scattering lobe, (2)~a Lommel-Seeliger-like (flat) disk function at small phase angles in the visible wavelength range, and (3)~monotonic reddening replacing the arch shape of the reflectance ratio for wavelength intervals associated with a positive spectral slope. The phase curve steepening can be understood in terms of the shadows cast by the fairy castles, whereas the flattening of the disk function results from the homogenization of the incidence and emergence angles over the surface, meaning that the three-dimensional structure of the fairy castles creates all possible combinations of these angles that the viewing geometry allows. It is not clear what causes the monotonic reddening, but it may also be related to the homogenization of the photometric angles. The fairy castles flatten the disk function over the visible wavelength range, preventing limb-darkening. Low-albedo planetary bodies are presumed to have a flat, Lommel-Seeliger-type disk function at low phase angle due to the predominance of single scattering, as predicted by a radiative transfer model \citep{H81}. Our results for basalt powder show how the disk function of a relatively high albedo surface is flat due to microscopic roughness, even though multiple scattering is clearly abundant. Thus, the absence of limb darkening does not necessarily imply that multiple scattering is negligible (e.g.\ \citealt{B96}). The bright sprinkled basalt powder surface is actually a good analog for Vesta, which basaltic surface has a similar visual albedo (0.38) and an equally flat disk function at low phase angle \citep{L13,S13}. If the Vesta regolith is composed of small particles, its surface is probably rough on a microscopic scale.

Could fairy castles be abundant in the solar system? They are composed of particles for which the Van der Waals force dominates over gravity. On Earth, these structures are on the order of 100~\textmu m, but still very large compared to the particle size. On small planetary bodies like asteroids they could be much larger because of the lower gravity and absence of air \citep{S10}. Particles that are small enough to build fairy castles should be abundant in the regolith of atmosphereless planetary bodies due to ``impact gardening''. However, they may not easily form. In experiments by \citet{SA67} fairy castles did not form when olivine particles were dropped from over 30~cm height under ultra-high vacuum conditions, for which particles are more adhesive (particle size $<44$~\textmu m, bulk 0-10~\textmu m). \citet{HS99} determined the roughness of the Lunar surface at microscopic scales from photographs of the Apollo Lunar Surface Closeup Camera. They found that roughness at sub-mm to sub-cm scales is well described by fractal statistics, with roughness increasing with decreasing size scale. While the soil sometimes appears clumpy, fairy castles are not evident in the images. However, the spatial resolution of the Apollo camera of 85~\textmu m may have been too low to detect fairy castles, which, in our experiment, were sized on the order of 100~\textmu m (although they could be larger on the Moon, given the lower gravity and absence of air). Thus, the fairy castles in our experiment merely represent a form of microscopic roughness, and one that may not be dominant in the solar system. Nevertheless, there is ample evidence that the regolith of atmosphereless planetary bodies is generally rough on a microscopic scale. Our experiments suggest that microscopic roughness is sufficient to explain the three photometric properties mentioned above, that are commonly observed for planetary bodies. Roughness on a macroscopic scale is not necessarily required.

Our ray-tracing model reproduces several key spectral changes that were observed in the experiments. How realistic is this geometric optics approach for describing the physics of scattering by a particulate surface? Light propagation in particulate media is a complex problem that has not yet been rigorously solved, especially for large irregular particles. In our experiments, the size ranges for the most of our samples are $>10$~\textmu m. In this case, only approximate methods are available, as computer requirements for existing accurate solutions dramatically increase for many-particle systems. Instead of widely used variations of the radiative transfer approach, like the \citet{H81} model, we use the ray-tracing technique of \citet{GS08}. The advantages of our approach are the explicit representation of the sample and the simple and direct connection of its physical parameters with the calculated optical properties. Our model reproduces key aspects of the observed spectral changes, like the arched shape of the reflectance ratio and monotonic reddening. Of course, representing the incident electromagnetic radiation as a set of independent rays is an approximation, that ignores potentially important wave effects. The particles in our experimental samples have fine structure on their surfaces and probably in their interiors. We hold these imperfections responsible for the monotonic reddening we observe in our experiments. Implementing particle surface roughness in our model by adopting Lambertian surface elements leads to monotonic reddening and repression of the reflectance ratio arch. However, we are not able to model the observed wavelength dependency. This is not surprising, as \citet{M07} already concluded that to accurately model surface roughness, we need something different from the Lambertian elements. In their effort to reproduce the scattering properties of Saharan dust particles in the geometric optics approach, these authors found that the single-scattering properties could not be accurately modeled without accounting for the effects of surface roughness. To our knowledge, a realistic implementation of small-scale surface roughness for irregular particles in the geometric optics approximation has not yet been developed, despite steps in the right direction \citep{M09,K12,LP13}.

Two other explanations for phase reddening have recently been put forward in the literature. First, \citet{H12} perform fits with a version of the \citet{H81} model to imaging data of the Lunar Reconnaissance Orbiter camera. They conclude that phase reddening is due to increased multiple scattering at large phase angles as the wavelength and albedo increase. While their conclusion is somewhat self-evident, our work demonstrates that the fundamental nature of phase reddening changes, depending on the surface microscopic roughness. Their model does not account for this (note that the ``macroscopic roughness'' parameter was not included). Also, the authors claim that the particle phase function must be virtually independent of wavelength because of their model fit results. This contradicts with our modeling approach: In case of a positive spectral slope, the particles become more transparent with wavelength, which allows light to penetrate the sample more deeply at intermediate phase angles. Second, \citet{S12} relate phase reddening to the explicit appearance of the wavelength in the equation for the absorption coefficient $\alpha = 4 \pi \kappa / \lambda$. However, this equation merely defines the absorption coefficient in terms of the imaginary part $\kappa$ of the complex index of refraction (e.g.\ \citealt{BH83}), and, unless one assumes $\kappa$ is constant with wavelength, it cannot be used as an explanation for phase reddening.

Our work aims to aid interpretation of observations of phase reddening of solar system bodies in terms of physical properties of the regolith. Observations by spacecraft have expanded the possibilities for study because of their ability to acquire observations at a wide variety of photometric angles, that are not possible to achieve from Earth. In recent years, orbiters around several solar system bodies have characterized spectral changes depending on phase angle in great detail. \citet{K13} uncovered a special type of spectral change for the Moon, the color opposition effect, a local minimum in the reflectance ratio curves at a small phase angle. This effect, unfortunately outside our experimental range, has been attributed to coherent backscattering \citep{S96}. It has also become apparent that the slope of the Lunar spectrum may not monotonously increase, but flatten out beyond certain phase angle, or even reach a maximum \citep{K07,Y11}. This behavior of the reflectance ratio reminds us of some of the profile shapes encountered in our experiments. Dawn observed phase reddening for Vesta, albeit very weak, which is consistent with the shallow slope of its visible spectrum \citep{L13,S14}. The pyroxene absorption bands are found to deepen, but only up to a certain phase angle \citep{L14}. For asteroid Eros, on the other hand, reddening of the visible spectrum was clearly observed by NEAR \citep{C02}. In addition, the 1~\textmu m absorption band increased with phase angle, but the 2~\textmu m absorption band did not. MESSENGER did not uncover any systematic reddening with increasing phase angle for Mercury \citep{D11}, consistent with ground-based observations.

We may now be able to explain the Viking~1 observations of Martian soil mentioned in the introduction \citep{G81}. The arched shape of the reflectance ratio in Fig.~\ref{fig:Viking} is reproduced by both our observations and simulations. However, illumination by the Martian sky may also induce spectral changes. While the \citet{G81} reflectance ratio was corrected for atmospheric illumination by measuring the brightness of shadows, \citet{T99} concluded that such a correction is insufficient. Then, if not due to illumination by the Martian sky, the arched shape of the soil reflectance ratio indicates that the surface in front of the lander was composed of semi-transparent particles, and was smooth on the scale of the typical particle size. The arch has not been reported for flybys of planetary bodies that spanned a wide range of phase angles. Our results indicate that monotonous phase reddening naturally follows from the presence of a microscopically rough regolith. That this type of reddening is commonly observed suggests that microscopic roughness is ubiquitous in the solar system.

\section*{Acknowledgements}

We are grateful to Lyuba Moroz for her help in procuring the samples and Ines B\"uttner for her assistance in preparing the samples. Fred Goesmann kindly allowed the first author to use his scanning electron microscope. Comments by reviewers J.R. Johnson and Y.~Shkuratov helped to improve the manuscript. S.E.S. thanks Stefano Mottola for fruitful discussions and reviewing the draft paper. Development of the PHIRE goniometer has been supported by the Swiss National Science Foundation.


\bibliography{phase_reddening}

\newpage
\clearpage

\begin{table}
\centering
\caption{Rock types and particle size classes used in our experiments. We refer to the $<10$~\textmu m size class of basalt as ``powder'' in the text. The samples were collected near towns in Germany (state name in brackets).}
\vspace{5mm}
\begin{tabular}{lll}
\hline
Rock type & Size range (\textmu m) & Origin \\
\hline
Basalt    & $<10$     & Winterberg \\
          & 200-350   & (Nordrhein- \\
          & 500-630   & Westfalen) \\
          & 800-1000  & \\
Granite   & 45-100    & Dohna \\
          & 500-630   & (Sachsen) \\
          & 1250-2000 & \\
Limestone & 10-32     & Freiberg \\
          & 40-63     & (Sachsen) \\
          & 63-80     & \\
\hline
\end{tabular}
\label{tab:samples}
\end{table}

\newpage
\clearpage

\end{document}